\documentclass[twocolumn,amsmath,amssymb,longbibliography,aps,prl,reprint,
superscriptaddress,]{revtex4-1}
\usepackage{epsfig}
\usepackage{graphicx}
\usepackage{wrapfig}
\usepackage[utf8]{inputenc}
\usepackage{amsmath}
\usepackage{amssymb}
\usepackage{multirow}
\usepackage{ulem}

\usepackage{hyperref}
\usepackage{color}

\usepackage{dsfont}
\usepackage{comment}
\usepackage{physics}
\usepackage{booktabs}
\newcommand{\red}{}

\definecolor{orange}{rgb}{1.0, 0.5, 0.0}

\begin{document}

\title{Odd-parity Magnetism Driven by Antiferromagnetic Exchange}

\author{Yue Yu}
\affiliation{Department of Physics, University of Wisconsin--Milwaukee, Milwaukee, Wisconsin 53201, USA} 

\author{Magnus B. Lyngby}
\affiliation{Niels Bohr Institute, University of Copenhagen, DK-2100 Copenhagen, Denmark}

\author{Tatsuya Shishidou} 
\affiliation{Department of Physics, University of Wisconsin--Milwaukee, Milwaukee, Wisconsin 53201, USA}
\date{\today}

\author{Mercè Roig}
\affiliation{Department of Physics, University of Wisconsin--Milwaukee, Milwaukee, Wisconsin 53201, USA} 
\affiliation{Niels Bohr Institute, University of Copenhagen, DK-2100 Copenhagen, Denmark}

\author{Andreas Kreisel}
\affiliation{Niels Bohr Institute, University of Copenhagen, DK-2100 Copenhagen, Denmark} 

\author{Michael Weinert} 
\affiliation{Department of Physics, University of Wisconsin--Milwaukee, Milwaukee, Wisconsin 53201, USA}

\author{Brian M. Andersen}
\affiliation{Niels Bohr Institute, University of Copenhagen, DK-2100 Copenhagen, Denmark} 

\author{Daniel F. Agterberg}
\affiliation{Department of Physics, University of Wisconsin--Milwaukee, Milwaukee, Wisconsin 53201, USA} 

\date{\today}


\begin{abstract}

Realizing odd-parity, time-reversal-preserving, non-relativistic spin splitting is a central goal for spintronics applications. We propose a group-theory-based microscopic framework to induce odd-parity spin splitting from coplanar antiferromagnetic (AFM) states without spin-orbit coupling (SOC). We develop phenomenological models for 421 conventional period-doubling AFM systems in non-symmorphic space groups and construct minimal microscopic models for 119 of these. We find that these AFM states can attain three possible competing ground states.  These ground states all break symmetries in addition to those broken by the usual AFM order. Specifically, they give rise to either odd-parity spin-splitting, nematic order, or scalar odd-parity order related to multiferroicity. Our microscopic theories reveal that the odd-parity spin-splitting energy scale is generically large and further reveal that the scalar odd-parity order gives a non-zero Berry curvature dipole without SOC. We identify 67 materials in the Magndata database for which our theory applies. We provide DFT calculations on Fe-based materials that reveal an $h$-wave spin splitting consistent with our symmetry arguments and apply our microscopic model to determine the non-relativistic Edelstein response for CeNiAsO.\end{abstract}

\maketitle

{\it Introduction:} Generating spin-split band structures without spin-orbit coupling has emerged as a central goal in quantum magnetism. This interest is driven by the discovery of altermagnets, a newly classified collinear compensated magnetic state that breaks time-reversal symmetry~\cite{Hayami:2019,Yuan:2020_Jul,Mazin:2021,Smejkal:2022_Sep,Smejkal:2022_Dec,Bhowal:2022_Dec}. Altermagnets feature a large non-relativistic even-parity spin-splitting \cite{Krempasky:2024} that enable  spin-transport \cite{Gonzalez-Hernandez:2021_Mar,Han:2024} and spin calitronics \cite{Cui:2023,Liu:2024}. Recently,  altermagnetism  has inspired generalizations to  coplanar magnetic states which allow large  {\it odd-parity time-reversal preserving} non-relativistic spin-splittings \cite{Anne_pwave2023,Brekke:2024,Chen:2024,Xiao:2024,Yi:2024}. These $p$-wave, or more generally, odd-parity magnets exhibit spin-splittings that are odd under a sign change of the momentum. They provide nonrelativistic variants of the familiar Rashba and Dresselhaus SOC and hence are likely to play an important role in spintronics \cite{Manchon:2015}.

Spontaneous odd-parity time-reversal preserving transitions in the spin sector have previously been considered as Pomeranchuk instabilities of a spherical Fermi surface \cite{Wu:2007,Kiselev:2017,Wu:2018}. Specifically, it was suggested that a $p$-wave spin splitting (or $p$-wave magnetism) spontaneously emerges from an $l=1$ orbital angular momentum Pomeranchuk instability in the spin channel \cite{Wu:2007}. However, it was later shown that this instability is forbidden due to local spin conservation \cite{Kiselev:2017}, rendering this mechanism for a non-relativistic $p$-wave spin-splitting ineffective. In the more recent analysis of $p$-wave magnetism \cite{Anne_pwave2023,Brekke:2024,Chen:2024,Xiao:2024,Yi:2024}, this no-go theorem is implicitly circumvented by inducing $p$-wave magnetism from coplanar antiferromagnets (AFMs). Since such $p$-wave magnetic order is a secondary order parameter, the energetic arguments of Ref.~\cite{Kiselev:2017}, which rely on $p$-magnetism being a primary order parameter, no longer apply. This opens a viable route to generating non-relativistic odd-parity magnetism.


Although significant progress has been made in applying spin-space group symmetry arguments to induce odd-parity time-reversal preserving magnetism \cite{Anne_pwave2023,Chen:2024,Xiao:2024,Yi:2024}, realistic microscopic models that allow for deeper insight into the properties and origins of this state are rare \cite{Brekke:2024}. Here we present a group-theory-based microscopic template for inducing odd-parity spin splittings from 421 AFMs and provide explicit microscopic theories for over 100 of these. This provides realistic microscopic models for $p$-wave, $f$-wave, and $h$-wave spin splittings.
We show that these spin splittings are naturally induced by coplanar AFM states when no SOC is present. Furthermore, we find that the AFM states we consider here generically exhibit some sort of translation invariant induced order.  Specifically, we find that either  non-relativistic odd-parity spin-splitting, nematicity, or scalar odd-parity order must emerge in the AFM state. The nematicity that emerges generalizes the AFM-driven nematic order discussed in Fe-based AFMs \cite{Bohmer:2018,Christensen:2019}.  The scalar odd-parity order that emerges appears in the commensurate AFM state of orthorhombic RMnO$_3$ rare earth manganites that exhibit improper ferroelectricity \cite{Kimura:2003,Hur:2004,Sergienko:2006,Cheong:2007}. 

In the following, we first develop a phenomenological theory for this class of AFM states, identify relevant material candidates in the Magndata database, and then turn to general symmetry-based microscopic theories.\\

{\it Induced odd-parity magnetism:} Here, two key ingredients are required to induce odd-parity spin splittings. First, the AFM state must belong to an irreducible representation (IR) that is at least two-dimensional (2D)  so that there are two magnetic degrees of freedom. Second, the inversion operator for this IR must be non-trivial. {\red As discussed in more detail below}, both ingredients are natural consequences of non-symmorphic space groups.

\begin{figure}[t]
\centering
\includegraphics[width=8cm]{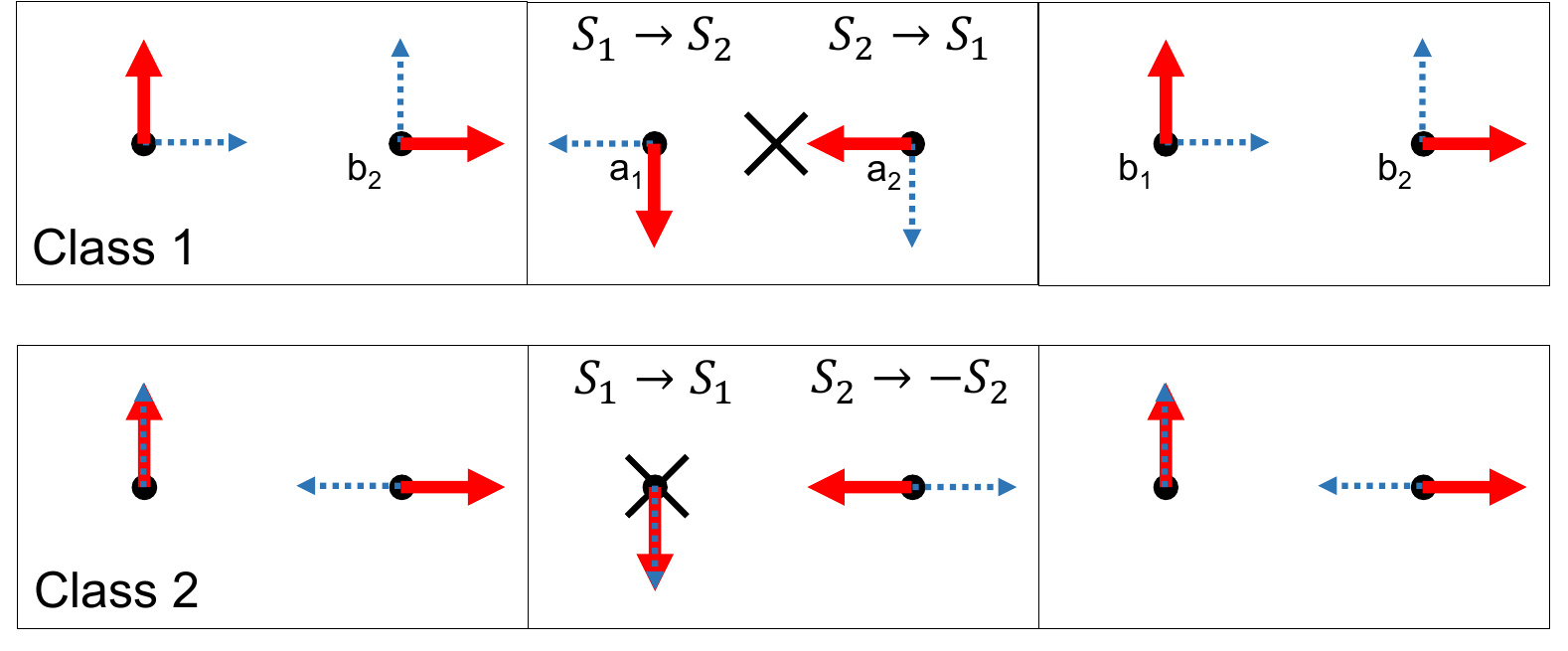}
\caption{Spin configuration of the coplanar AFM state, denoted by red solid arrows. Its inversion pair is illustrated by blue dashed arrows. The inversion center is at the X mark. Black solid lines denote the nonmagnetic unit cell. (Top) For sites related by inversion. (Bottom) For sites at the inversion center.
}
\label{F:illus}
\end{figure}

These two symmetry ingredients are illustrated in the simplest unit cell doubling AFM in Fig.~\ref{F:illus}. The non-
magnetic unit cell consists of two sublattice atoms. The
two sublattice magnetic moments, here labeled ${\bf S_1}$ and ${\bf S_2}$, typically form a 2D IR for AFM \cite{elcoro:2017}. As described below, consideration of the inversion operator naturally leads to two classes. In the class 1 (top panel), the two atoms are related by inversion, whereas in the class 2 (bottom panel), each atom sits at an inversion center. 

Here ${\bf S_1} \times {\bf S_2}$ is an important quantity, as we discuss later, it generates the odd-parity spin-splitting. This induced order is time-reversal invariant, as time-reversal symmetry flips both spins, and is translationally invariant. {\red Here, the inversion symmetry in non-symmorphic space groups acts non-trivially in the 2D IR \cite{elcoro:2017}.} In class 1, inversion interchanges ${\bf S_1}$ and ${\bf S_2}$. In class 2, inversion leaves ${\bf S_1}$ unchanged but flips ${\bf S_2}$. In both cases, the resulting spin splitting ${\bf S_1} \times {\bf S_2}$ is odd under inversion. Here ${\bf S_1} \times {\bf S_2}$ can also be interpreted as a non-relativistic version of the Dzyaloshinsky-Moriya vector \cite{Dzyaloshinsky:1958,Moriya:1960}. {\red In contrast, inversion symmetry within symmorphic space groups is always a trivial identity operation for any IR because it commutes with all other symmetry operations. Realizing odd-parity ${\bf S_1} \times {\bf S_2}$ then requires multiple IRs/phase transitions (See examples in End Matter).}

It is possible to identify all space groups and AFM ordering wavevectors that give rise to the two required symmetry ingredients. The nontrivial inversion operator requires non-symmorphic space groups, where inversion does not commute with all symmetries at certain time-reversal and inversion invariant momenta (TRIM) \cite{Szabo:2024}. This includes 325 IRs for $p$-wave, 84 IRs for $f$-wave, and 12 IRs for $h$-wave spin splittings. The list of space groups and wavevectors is in supplementary materials \cite{supp} (including Ref. \cite{FLAPW2009,KH1977,MPK1980,PBE,pearson1997handbook,TS2018,Koepernik1999}). In these 421 distinct scenarios, the AFM state with odd-parity magnetism can be stabilized through a single continuous phase transition. These 2D IRs are not restricted to cases with only two atoms per nonmagnetic unit cell. For systems with more atoms, the theory remains applicable. However, $\bf S_{1}$ (and $\bf S_{2}$) now describes spin order across multiple atoms in the non-magnetic unit cell. The $p$-, $f$-, and $h$-wave spin splittings are identified by analyzing other symmetry operators at these TRIMs.

\begin{figure}[htb]
\centering
\includegraphics[width=9cm]{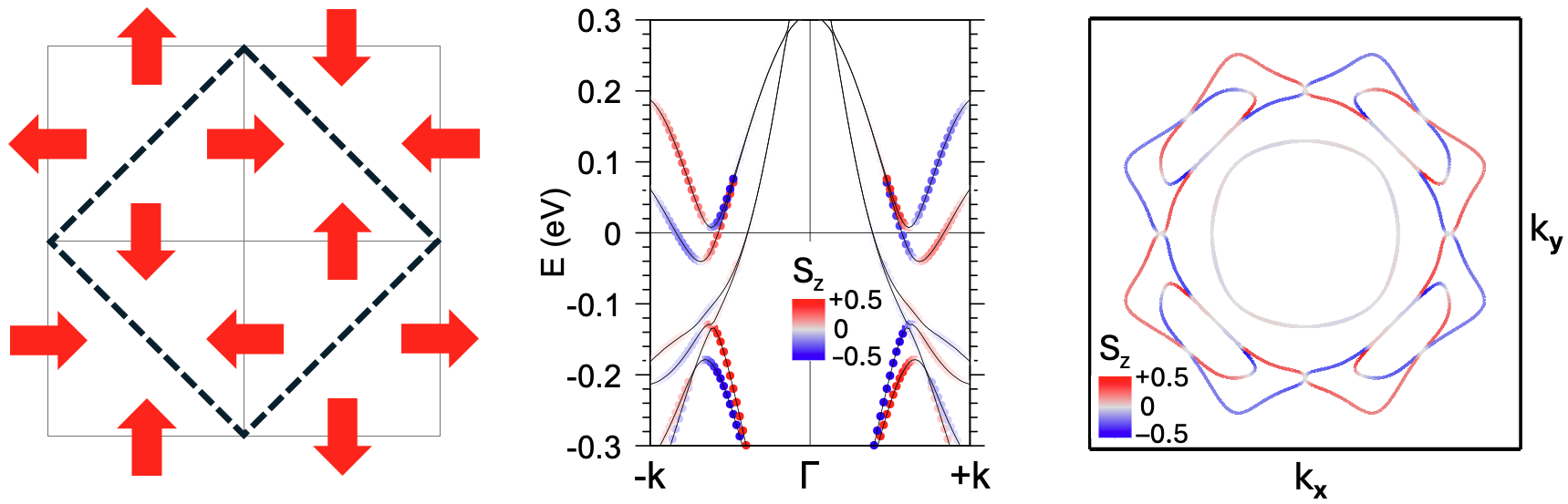}
\caption{(Left) FeSe coplanar magnetic state (Center) DFT band dispersion along the momentum cut $\pm (\pi,0.4\pi,0.5\pi)$. (Right) Fermi surface ($k_z =0.4\pi$ section) exhibiting $h$-wave spin splitting. The band spin induced is represented by a red-white-blue color code.}
\label{F:FeSe}
\end{figure}

To test our symmetry-based arguments, we have examined if the predicted $h$-wave $k_xk_yk_z(k_x^2-k_y^2)\sigma_z$ spin-splitting appears within a DFT calculation for FeSe with coplanar order at the M-point $(\pi,\pi, 0)$.  Collinear AFM ordering at the M point is known to generate the nematic order observed in many Fe-based materials \cite{Bohmer:2018}. A coplanar magnetic order is sometimes observed in this class of materials \cite{Ding:2017,Chmaissem:2022}. In particular, this coplanar state appears in LaOFeAs$_{1-x}$P$_x$ \cite{Chmaissem:2022} with space group 129, providing a materials  realization of this state. As shown in Fig.~\ref{F:FeSe}, the DFT results for this coplanar state agree with our symmetry prediction. Furthermore, this calculation reveals that the spin-splitting, even for a highly nodal $h$-wave state, can be as large as 0.1 eV.

\begin{figure}[h]
\centering
\includegraphics[width=8cm]{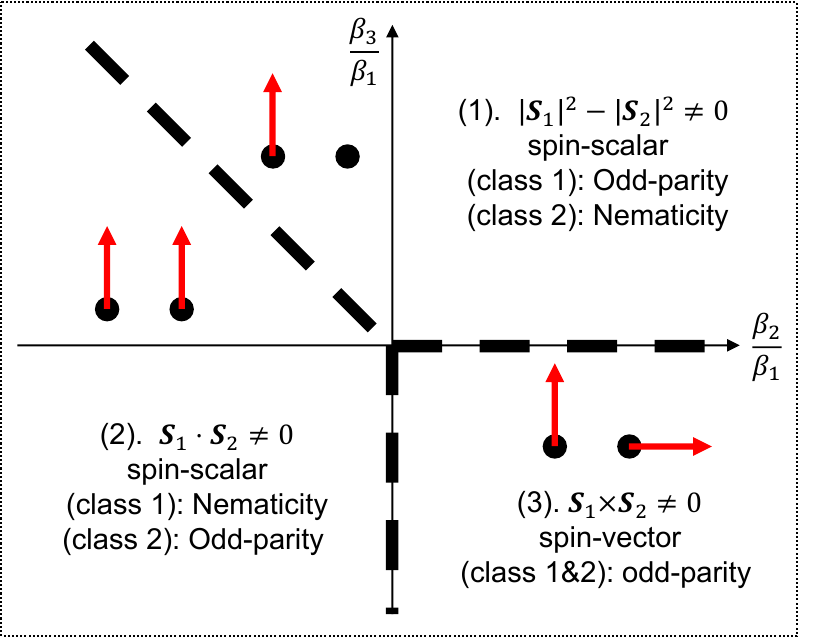}
\caption{Three competing phases and their induced orders. Phase boundaries are at $\beta_3=0$, $\beta_2=0$, and $\beta_2+\beta_3=0$.
}
\label{F:phase}
\end{figure}

\noindent {\it Phenomenological model:}
We apply Landau theory to describe the 421 different AFM phase transitions and analyze all the possible induced orders. In the absence of SOC, the free energy density of the 2D IR is \cite{fernandes:2016}
\begin{equation}
\begin{split}
f&=\alpha(T)\sum_i{\bf S}_i\cdot {\bf S}_i+\beta_1(\sum_i{\bf S}_i\cdot {\bf S}_i)^2\\&+\beta_2({\bf S}_1\cdot {\bf S}_2)^2+\beta_3({\bf S}_1\cdot {\bf S}_1)({\bf S}_2\cdot {\bf S}_2).
\end{split}
\end{equation}
Here, we have included all the symmetry-allowed terms to quartic order in $\bf S_{1,2}$. {\red In $f$, spin rotational invariance allow terms ${\bf S}_1\cdot {\bf S}_2$ and $|{\bf S}_1|^2-|{\bf S}_2|^2$ to also appear, here non-symmorphic symmetries are important since they enforce that these terms are not permitted.} The continuous phase transition occurs as $\alpha(T)\rightarrow0$. Depending on $\beta_i$, there are three competing AFM states with different uniform induced orders (Fig.~\ref{F:phase}).

Among these induced orders, the spin-vector ${\bf S}_1\times{\bf S}_2$ exhibits odd-parity spin splitting, while the other two are spin-scalars, whose characteristics differ depending on the site symmetry. In the class 1, $(|{\bf S}_1|^2-|{\bf S}_2|^2)$ is odd-parity, while $({\bf S}_1\cdot{\bf S}_2)$ is even-parity (nematicity). In the class 2, their physical meanings are opposite. 
This phenomenological model describes 45 odd-parity states (14 spin-vector, 24 spin-scalar, and 7 mixed by SOC) and 22 nematic materials in the Magndata database \cite{Gallego:2016} (list included in SM). 

The odd-parity spin order ${\bf S_1} \times {\bf S_2}$, as a spin vector, can generate non-relativistic spin-related odd-parity responses, including odd-parity spin-splitting and the non-relativistic Edelstein effect. On the other hand, the scalar odd-parity orders, which are invariant under spin rotation, produce non-relativistic spin-independent odd-parity  responses, including nonlinear I-V transport and ferroelectric-like order. To estimate the magnitudes of the induced orders and to evaluate the response properties of the magnetic states, we develop symmetry-based minimal microscopic models.\\

\noindent {\it Microscopic models:}
For pedagogical reasons, we first introduce a simplified model. Specifically, we consider the class 1 shown in Fig.~\ref{F:illus} and keep only NN hopping $t$ and the magnetic order $\vec{J}_i$
\begin{equation}\label{eq:pedagogical}
\begin{split}
H&=\begin{pmatrix}
h_0({\bf k})&  O_M\\ O_M^\dagger&\widetilde{h_0}({\bf k+Q})
\end{pmatrix}\\&=\begin{pmatrix}
\epsilon_k+t\cos \frac{k_x}{2} \tau_x&  O_M\\ O_M^\dagger&\epsilon_{k+Q}+t\sin \frac{k_x}{2}\tau_y
\end{pmatrix}
\end{split}
\end{equation}
Here $\tau_i$ are Pauli matrices for nonmagnetic sublattices. The on-site exchange driven spin ordering is described by $O_M=\vec{J}_1\cdot\vec{\sigma}+\tau_z\vec{J}_2\cdot\vec{\sigma}$. As in the usual SDW theory\cite{auerbach:2012}, $H$ features sublattice-independent hopping $\epsilon_k$ and $\epsilon_{k+Q}$ on the diagonal blocks. The model also exhibits inter-sublattice hopping $\tau_{x,y}$. {\red The differences between $h_0({\bf k})$ and $\widetilde{h_0}({\bf k+Q})$ in Eq.~(\ref{eq:pedagogical}) arise from the {\bf Q}-originated Bloch phase: In $h_0({\bf k})$, the hopping from $(a_2,b_2)$ to $a_1$ is real, $t/2(e^{ik_x/2}+e^{-ik_x/2})=t\cos\frac{k_x}{2}$. In $\widetilde{h_0}({\bf k+Q})$, however, the relative phase factor of $-1$ between the $a$ and $b$ cells arising from $\mathbf{Q}$ causes this hopping to be imaginary, $t/2(-e^{ik_x/2}+e^{-ik_x/2})=-it\sin\frac{k_x}{2}$, and then to account for the resulting factor of $i$, $\tau_x$ is transformed into $\tau_y$.} The transformation from $\tau_x$ to $\tau_y$ can also be understood as a consequence of time-reversal symmetry that exists in the nonmagnetic state: the odd time-reversal symmetry of $\sin \frac{k_x}{2}$ can only be compensated by transforming $\tau_x$ to $\tau_y$.


The $8\times8$ pedagogical Hamiltonian Eq.~\eqref{eq:pedagogical} captures the essence of nonsymmorphic crystal symmetries with the two sublattices, translational symmetry breaking with unit cell doubling, and spin orderings.
We now extend the model to general tight-binding models with Wyckoff position of multiplicity two, including all symmetry-allowed hopping for different nonsymmorphic space groups. This minimal model captures 13 nonsymmorphic materials (listed in SM) with odd-parity AFM. The models for the two inversion classes differ slightly. We explicitly consider the class 1 below and include the models for the class 2 in the SM. The Hamiltonian is
\begin{equation}
\begin{split}
&H({\bf k})=\begin{pmatrix}
h_0({\bf k})&O_M\\O_M^\dagger&\widetilde{h_0}({\bf k+Q})
\end{pmatrix}\\&=\begin{pmatrix}
\epsilon_0+t_0+t_x\tau_x+t_y\tau_y &  \vec{J}_1\cdot\vec{\sigma}+\tau_z\vec{J}_2\cdot\vec{\sigma}\\ \vec{J}_1\cdot\vec{\sigma}+\tau_z\vec{J}_2\cdot\vec{\sigma}&\epsilon_0-t_0+\widetilde{t_x}\tau_x+\widetilde{t_y}\tau_y
\end{pmatrix}\\
&\equiv\epsilon_0+t_1\rho_0\tau_x+t_2\rho_0\tau_y+t_3\rho_z\tau_x+t_4\rho_z\tau_y\\&+t_0\rho_z+\rho_x(\vec{J}_1\cdot\vec{\sigma}+\tau_z\vec{J}_2\cdot\vec{\sigma}),
\end{split}
\label{E:3}
\end{equation}
where $t_1 \equiv (t_x + \widetilde{t_x})/2$, $t_2 \equiv (t_y + \widetilde{t_y})/2$, $t_3 \equiv (t_x - \widetilde{t_x})/2$, and $t_4 \equiv (t_y - \widetilde{t_y})/2$. The Pauli matrices $\tau_i$ and $\rho_i$ relate the two nonmagnetic sublattices (1,2) and {\red states at ($\mathbf{k}$,$\mathbf{k+Q}$) in the folded Brillouin zone}, respectively. The time-reversal operator and inversion operator are $T=i\sigma_yK$ and $I=\tau_x$, with momentum flip.
The first block, $h_0({\bf k})$ is the same as the nonmagnetic Hamiltonian, containing k-even $\tau_x$ and k-odd $\tau_y$ hopping terms, whose detailed form differs in different nonsymmorphic space groups (listed in SM). $\widetilde{h_0}({\bf k+Q})$ is obtained from  $h_0({\bf k+Q})$ by correcting $\tau_{x,y}$ considering time-reversal symmetry {\red and the Bloch phase factor}.
The $\epsilon_0$ (or $t_0$) term includes the even (or odd) intra-sublattice hopping terms under $\bf k\rightarrow k+Q$, such as $\cos k_y$. {\red The pedagogical model in Eq.~\ref{eq:pedagogical} results by taking $t_y=\tilde{t}_x=0$, with $\epsilon_k=\epsilon_0+t_0$, $\epsilon_{k+q}=\epsilon_0-t_0$, $t_x=t\cos k_x/2$, and $\tilde{t}_y=t\sin k_x/2$.} As a next-nearest neighbor hopping, $t_0$ term does not significantly affect the magnitude of the odd-parity spin splitting, and the subsequent analysis assumes $t_0=0$.
The three AFM states are described by $(\vec{J}_1,\vec{J}_2)=(\frac{{\bf S_1+S_2}}{2},\frac{{\bf S_1-S_2}}{2})=J(\hat{x},\hat{x}),J(\hat{x},0),\text{and }J(\hat{x},\hat{y})$. {\red The real-space interpretation of Eq.\ref{E:3} can be found in End Matter.}

{\it Non-relativistic odd-parity spin-splittings:}
The odd-parity spin splitting can be obtained from the dispersion:
\begin{equation}
\begin{split}
&E_{\alpha\beta\gamma}=\epsilon_0+\gamma\bigl\{\vec{J}_1^2+\vec{J}_2^2+|{\bf t}|^2+2\beta\bigl[(t_1t_3+t_2t_4)^2\\&+(\vec{J}_1^2+\vec{J}_2^2)|{\bf t}|^2/2+(\vec{J}_1^2-\vec{J}_2^2)(t_1^2+t_2^2-t_3^2-t_4^2)/2\\&+ (\vec{J}_1\cdot\vec{J}_2)^2+2\alpha (t_1t_4-t_2t_3)|\vec{J}_1\times\vec{J}_2|\bigr]^{1/2}\bigr\}^{1/2}, 
\end{split}
\label{E:dispersion}
\end{equation}
where $\alpha,\beta,\gamma=\pm$ and $|{\bf t}|^2\equiv t_1^2+t_2^2+t_3^2+t_4^2$. The spin splitting is between $\alpha=\pm$ bands.
For the first two AFM states, $E_{\alpha\beta\gamma}$ is independent of $\alpha$, implying doubly degenerate bands. For the coplanar state, the magnitude of spin splitting depends on $J\sqrt{t_1t_4-t_2t_3}\propto J(t_x\widetilde{t_y}-\widetilde{t_x}t_y)^{1/2}$. These spin splittings for different space groups and AFM wavevectors are shown in SM.

We now consider SG129(P4/nmm) with Wyckoff position 2a, relevant for FeSe (and 2c, for CeNiAsO \cite{Anne_pwave2023}). The coefficients in the Hamiltonian are:
$t_x=t_{x0}\cos\frac{k_x}{2}\cos\frac{k_y}{2}$ and $t_y=t_{y0}\cos\frac{k_x}{2}\cos\frac{k_y}{2}(\cos k_x-\cos k_y)\sin k_z
$ (for 2c, remove $\cos k_x-\cos k_y$ factors here and below). 
For the AFM state with ${\bf Q}=(\pi,0,0)$, the other coefficients are
$
\widetilde{t_x}=-t_{y0}\sin\frac{k_x}{2}\cos\frac{k_y}{2}({\red-}\cos k_x-\cos k_y)\sin k_z$ and $\widetilde{t_y}=t_{x0}\sin\frac{k_x}{2}\cos\frac{k_y}{2}$, from which we get $p$-wave spin splitting proportional to $\sin k_x$. For the AFM state at $(\pi,\pi,0)$, the other coefficients are
$\widetilde{t_x}=t_{x0}\sin\frac{k_x}{2}\sin\frac{k_y}{2}$ and $\widetilde{t_y}=-t_{y0}\sin\frac{k_x}{2}\sin\frac{k_y}{2}(\cos k_x-\cos k_y)\sin k_z$,
from which we obtain an $h$-wave spin splitting proportional to $\sin k_x\sin k_y(\cos k_x-\cos k_y)\sin k_z$ found in Fig.~\ref{F:FeSe}. 

\begin{figure}[h]
\centering
\includegraphics[width=\linewidth]{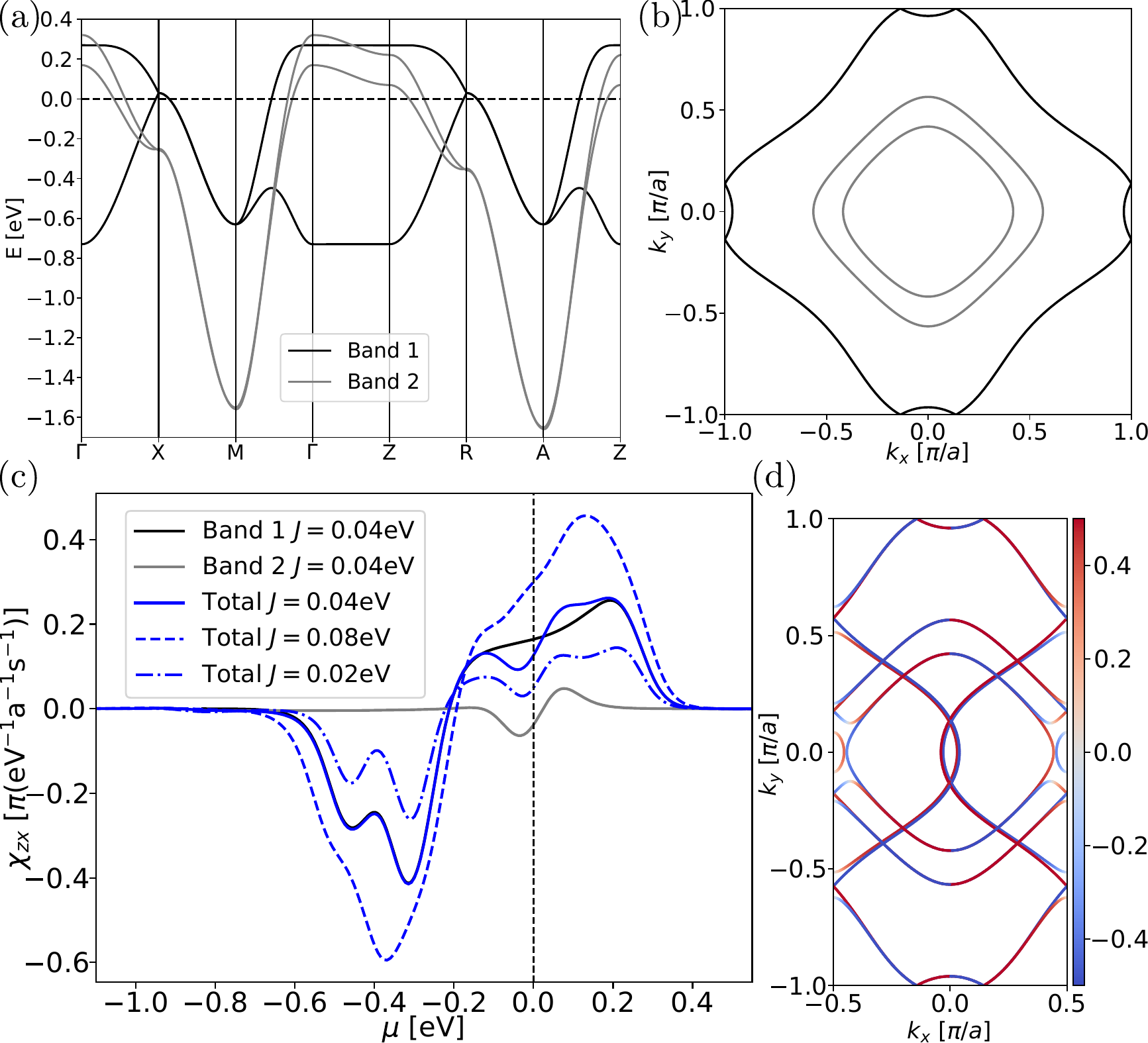}
\caption{(a) Band structure relevant for CeNiAsO with two pairs of single-orbital models (black/gray). (b) Corresponding Fermi surface cut at $k_z=0$ in the BZ of the nonmagnetic system. (c) Edelstein susceptibility $\chi_{zx}$ as a function of chemical potential $\mu$ for different $p$-wave order parameters $J$, displaying the contributions of each band. (d) Fermi surface with the expectation value of the spin operator in the magnetic state exhibiting odd-parity spin splitting.}
\label{fig:CeNiAsO}
\end{figure}

{\it Non-relativistic Edelstein Effect:} The Edelstein effect refers to the accumulation of spin-density induced by electric fields/currents, and serves as an important mechanism for charge-to-spin conversion. Traditionally, the Edelstein effect is relevant for systems with strong SOC, but has recently been proposed as a characteristic feature of certain non-collinear magnetic structures. This heralds the concept of a non-relativistic Edelstein effect~\cite{Gonzalez2024,Hu2024,Chakraborty2024}. Time-reversal symmetric odd-parity magnets are prime examples featuring the non-relativistic Edelstein effect. To demonstrate this for a specific case, we focus on the $p$-wave material candidate CeNiAsO~\cite{Anne_pwave2023,Hu2024}. In Fig.~\ref{fig:CeNiAsO}(a,b) we show the minimal band structure and the associated normal state Fermi surface obtained from tight-binding fits to the low-energy bands from DFT calculations without SOC, see SM. The spin-polarization of the bands in the $p$-wave magnetic state is shown in Fig.~\ref{fig:CeNiAsO}(d).

We compute the Edelstein response tensor using the Boltzmann equation in the relaxation-time approximation, going to linear order in the electric field. This gives the transport coefficient $\chi_{ij} =\chi_0 \sum_\nu \int d^2k \ (-\partial_{\epsilon} f_0) \langle \sigma_{i\nu} \rangle(k) v_j$, where $\chi_0 = \frac{\tau |e| \mu_B \mathcal{A}}{4\pi^2}$, $\nu$ is the band index, $v_i = \partial_{k_i} \epsilon(k)$, $f_0$ is the Fermi-Dirac distribution function and $\langle \sigma_{j\nu} \rangle(k)$ is the expectation value of the Pauli spin matrix in direction $j$. For $\chi_0$, $\tau$ denotes the relaxation time, $\mu_B$ is the Bohr magneton and $\mathcal{A} = a_x a_y$ is the area of the unit cell. Calculating this response for the bands shown in Fig.~\ref{fig:CeNiAsO}(c) gives $\chi_{zx}$ displayed in Fig.~\ref{fig:CeNiAsO}(c). As seen, most of the non-relativistic Edelstein response originates from the first band (black line), and we observe that the response changes sign as a function of the chemical potential~\cite{Hu2024}.

{\it Nematic Order:}
Our microscopic model also reveals nematic order in the electronic spectrum. This appears in Eq.~\eqref{E:dispersion} as the coefficient of $|\vec{J}_1|^2-|\vec{J}_2|^2$, which is $(t_1^2+t_2^2)-(t_3^2+t_4^2)=t_x\widetilde{t_x}+t_y\widetilde{t_y}$. Specifically, we consider the well-known example of Fe-based superconductors with SG129 with ${\bf Q}=(\pi,\pi,0)$. Here $t_x=t_{x0}\cos\frac{k_x}{2}\cos\frac{k_y}{2}$ so that  $\widetilde{t_x}=t_{x0}\sin\frac{k_x}{2}\sin\frac{k_y}{2}$. This yields a $\sqrt{Jt_{x0}}\sin k_x \sin k_y$-like nematic term in the dispersion. In the SM, we identify 22 materials in the Magndata database that should exhibit a similar nematic-driven electronic anisotropy.  

{\it Spin-scalar odd-parity state:} We now consider the scalar odd-parity state. We will focus on state 1 in class 1 below, and the analysis for state 2 in class 2 is provided in SM. Although the state $(\vec{J}_1,\vec{J}_2)=J(\hat{x},\hat{x})$ has no spin-splitting [see Eq.~\eqref{E:dispersion}] and does not exhibit any symmetry breaking in the electronic dispersion, it can generate a non-relativistic Berry curvature dipole (unlike the spin-vector or nematic states), strongly contributing to nonlinear I-V transport. The Berry curvature is:
\begin{equation}
\begin{split}
\Omega^{ij}_{\alpha\beta\gamma}&=\frac{\beta}{|\vec{h}|^3}\sum_{abc}\epsilon_{abc}[h_a\partial_{k_i}h_b\partial_{k_j}h_c+\widetilde{h}_a\partial_{k_i}\widetilde{h}_b\partial_{k_j}\widetilde{h}_c],
\\\vec{h}&\equiv(Jt_x,Jt_y,J^2+(-t_x^2-t_y^2+\widetilde{t}_x^2+\widetilde{t}_y^2)/4),\\\vec{\widetilde{h}}&\equiv(J\widetilde{t}_x,J\widetilde{t}_y,J^2+(+t_x^2+t_y^2-\widetilde{t}_x^2-\widetilde{t}_y^2)/4).
\end{split}
\end{equation}
The derivation is in the SM. $\epsilon_{abc}$ is the antisymmetric tensor. The nontrivial k-odd dependence from $t_xt_y$ carries the site symmetry, as the AFM state distinguishes the two sublattices (state 1 in Fig.~\ref{F:phase}). This k-odd Berry curvature can generate non-zero Berry curvature dipoles $\langle d\Omega^{ij}/dk_m\rangle_{FS}$, which drives non-relativistic nonlinear I-V transport coefficients \cite{Sodemann:2015,Kaplan:2024,watanabe:2024}. In addition, this state is related to multiferroicity \cite{Cheong:2007,matsuda:2024} and will also exhibit ferroelectric-like order\cite{Cheong:2007}.


{\it Conclusions:} We have developed a group-theory-based framework for generating odd-parity, time-reversal-preserving spin splitting from coplanar AFM states without SOC. By analyzing phenomenological models of 421 period-doubling AFM systems in nonsymmorphic space groups, we constructed minimal microscopic models for 119 of them, identifying three competing ground states: odd-parity spin splitting, nematic order, and scalar odd-parity order. Our DFT calculations demonstrate robust $h$-wave spin splitting in FeSe, characterized by a significant spin-splitting energy. Additionally, our minimal microscopic model reveals the non-relativistic Edelstein response in odd-parity spin-splitting states and non-relativistic nonlinear transport in scalar odd-parity states. We focused on applications to CeNiAsO and FeSe, with the model applicable to 67 AFM materials in Magndata database, providing a broad framework for spintronics application in realistic systems.

{\it Acknowledgments} 
We thank Philip Brydon, Rafael Fernandes, Ian Fisher, and Changhee Lee for useful discussions. 
M.~R. acknow\-ledges support from the Novo Nordisk Foundation grant NNF20OC0060019. A.~K. acknowledges support by the Danish National Committee for Research Infrastructure (NUFI) through the ESS-Lighthouse Q-MAT. D.~F.~A. and Y.~Y. were supported by the U.S. Department of Energy, Office of Basic Energy Sciences, Division of Materials Sciences and Engineering under Award No. DE-SC0021971 for symmetry-based calculations. M.~W. and T.~S. were supported by National Science Foundation Grant No. DMREF 2323857 for DFT calculations.

\bibliography{Odd_parity}

\onecolumngrid
\section{End Matter}
\twocolumngrid
The data that support the findings of this article are openly available\cite{dataset}.

\subsection{Examples on non-symmorphic symmetries}
{\red
We now highlight the importance of non-symmorphic symmetries through an example in space group 129. The results for other space groups can be found in Table.I in SI. We consider wavevector at X-point $(\pi,0,0)$, relevant for CeNiAsO. At X-point, the real-space symmetries keeping momentum invariant are inversion P, mirror symmetries $\widetilde{M_x}=\{M_x|\frac{1}{2},0,0\}$, $\widetilde{M_y}=\{M_y|0,\frac{1}{2},0\}$, $\widetilde{M_z}=\{M_z|\frac{1}{2},\frac{1}{2},0\}$, and their products. With inversion center chosen at the origin, other non-symmorphic symmetries carry a translational part, such as $\{\frac{1}{2},0,0\}$ in $\widetilde{M_x}:(x,y,z)\rightarrow(-x+1/2,y,z)$. 

Due to the translational part, inversion symmetry P does not commute with $\widetilde{M_x}$: 
\begin{equation}
\begin{split}
\widetilde{M_x}P:(x,y,z)\rightarrow(-x,-y,-z)\rightarrow(x+\frac{1}{2},-y,-z)\\P\widetilde{M_x}:(x,y,z)\rightarrow(-x+\frac{1}{2},y,z)\rightarrow(x-\frac{1}{2},-y,-z)
\end{split}
\end{equation}
The difference is a translation ${\bf r}=\{1,0,0\}$, which is equal to $\exp(i{\bf k \cdot r})=-1$ at the X-point, so the two symmetries anticommute. Consequently, the irreducible representation (IR) consists of two-dimensional Pauli matrices $\tau_i$, and inversion is non-trivial\cite{elcoro:2017}: $P=\tau_x, \widetilde{M_x}=\tau_z, \widetilde{M_y}=\tau_0,\widetilde{M_z}=i\tau_y$. 
In this 2D IR for AFM, since $({\bf S}_1,{\bf S}_2)$ transforms to $({\bf S}_2,{\bf S}_1)$ under $P=\tau_x$, ${\bf S}_1\times {\bf S}_2$ is odd-parity. Similarly, ${\bf S}_1\times {\bf S}_2$ is odd under $\widetilde{M_x}$, while even under $\widetilde{M_{y,z}}$. The spin splitting is then proportional to $k_x$.

In contrast, symmetries in symmorphic space groups do not have the translational part. Consequently, inversion symmetry commute with other symmetries, and must be represented as identity matrix in any IR. To have odd-parity ${\bf S}_1\times {\bf S}_2$, multiple IRs and multiple phase transitions are needed.

\subsection{Real-space interpretation of Eq.~\ref{E:3}}
We note that Eq.\ref{E:3} can also be derived from a tight-binding model of four sublattices, with the Pauli matrices $\tau_i$ and $\widetilde\rho_i$ acting on nonmagnetic (1,2) sublattices and neighboring ($a$,$b$) unit cells, respectively, i.e., explicitly starting from a supercell representation. For the non-magnetic Hamiltonian matrix, the $a$-$a$ and $b$-$b$ blocks will be identical, thus transforming as $\widetilde\rho_0$, while the off-diagonal hopping terms connecting the ($a$,$b$) unit cells are proportional to $\widetilde\rho_x$ (or more generally $\cos\alpha\,\widetilde\rho_x + \sin\alpha\,\widetilde\rho_y$). In this representation the AFM states alternate between the $a$ and $b$ blocks, and hence are represented by $\widetilde\rho_z$. The unitary transformation $(\widetilde\rho_x + \widetilde\rho_z)/\sqrt{2}$ will block-diagonalize the non-magnetic Hamiltonian -- with a change of basis: $\widetilde\rho_x \to \rho_z$ and $\widetilde\rho_z \to \rho_x$ -- leading to the Hamiltonian of Eq.~\ref{E:3}.  This unit-cell-doubling coplanar and collinear AFM state hosts the symmetry $[\rho_z\sigma_z,H]=0$, which further block-diagonalizes the Hamiltonian and allows for analytical results. In real space, after back-transforming to the original $\widetilde\rho_i$ basis, this symmetry is a two-fold spin rotation about the $z$-axis ($\sigma_z$) followed by one-lattice-spacing translation that interchanges the ($a$,$b$) unit cells ($\widetilde\rho_x$).
}

\clearpage
\appendix
\onecolumngrid
\section{Supplementary Material}
\section{List of tables}
Table.\ref{T:1} contains the symmetry of the odd-parity spin splitting, enforced by the nonsymmorphic space group symmetries.

Table.\ref{T:2} contains the symmetry of the nematicity, enforced by the nonsymmorphic space group symmetries.

Table.\ref{T:3} contains the symmetry of the scalar odd-parity order, enforced by the nonsymmorphic space group symmetries.

Table.\ref{T:4} contains the microscopic tight-binding parameters and the resulting spin splittings, for the class 1, where two atoms are related by inversion. 

Table.\ref{T:5} contains the microscopic tight-binding parameters and the resulting spin splittings, for the class 2, where two atoms are at inversion center. 

\section{Class 1}
Here, we provide the detailed step to get the dispersion and Berry curvature for class 1, where the atoms related by inversion. The Hamiltonian is:
\begin{equation}
	\begin{split}
		H=t_1\rho_0\tau_x+t_2\rho_0\tau_y+t_3\rho_z\tau_x+t_4\rho_z\tau_y+\rho_x\vec{J_1}\cdot\vec{\sigma}+\rho_x\tau_z\vec{J_2}\cdot\vec{\sigma},
	\end{split}
\end{equation}
where $t_1 \equiv (t_x + \widetilde{t_x})/2$, $t_2 \equiv (t_y + \widetilde{t_y})/2$, $t_3 \equiv (t_x - \widetilde{t_x})/2$, and $t_4 \equiv (t_y - \widetilde{t_y})/2$. We have taken $t_0=0$ and dropped the trivial $\epsilon_0$ term for simplicity. 

The three AFM states are described by $(\vec{J}_1,\vec{J}_2)=(\frac{{\bf S_1+S_2}}{2},\frac{{\bf S_1-S_2}}{2})=J(\hat{x},\hat{x}),J(\hat{x},0),\text{and }J(\hat{x},\hat{y})$. Without loss of generality, we will consider $\vec{J}_{1,2}$ in the xy plane. Using $[\rho_z\sigma_z,H]=0$, we can block-diagonalize the Hamiltonian:
\begin{equation}
	\begin{split}
		&u^\dagger Hu=\left(\begin{array}{cc}
			H_{\alpha=+}&  \\
			&H_{\alpha=-} 
		\end{array}\right),\;u=\left(\begin{array}{cccc}
			1&&&\\&&1&\\&&&1\\&1&&
		\end{array}\right),\\&H_\alpha=t_1\widetilde{\rho_0}\tau_x+t_2\widetilde{\rho_0}\tau_y+t_3\widetilde{\rho_z}\tau_x+t_4\widetilde{\rho_z}\tau_y+\vec{J}_{1\alpha}\cdot\vec{\widetilde{\rho}}+\tau_z\vec{J}_{2\alpha}\cdot\vec{\widetilde{\rho}},
	\end{split}
\end{equation}
where the unitary transformation $u$ is in $(\rho,\sigma)$ space, and $\vec{J}_{i\alpha}\equiv(J_{ix},\alpha J_{iy},0)$. Using $\{\widetilde{\rho_z}\tau_z,H_\alpha\}=0$, we can further off-block-diagonalize the resulting $4\times4$ Hamiltonian:
\begin{equation}
	\begin{split}
		\widetilde{H}_\alpha&=\widetilde{u}^\dagger H_\alpha \widetilde{u}=\left(\begin{array}{cc}
			0&M\\ M^\dagger &0
		\end{array}\right), \;\widetilde{u}=\left(\begin{array}{cccc}
			1&&&\\&&1&\\&&&1\\&1&&
		\end{array}\right),\\M&\equiv\left(\begin{array}{cc}t_1-it_2+t_3-it_4&
			J_{1x}+J_{2x}-i\alpha J_{1y}-i\alpha J_{2y}\\J_{1x}-J_{2x}+i\alpha J_{1y}-i\alpha J_{2y}&t_1+it_2-t_3-it_4
		\end{array}\right), 
	\end{split}
\end{equation}
The eigenvectors $[v_1, v_2]^T$ of $\widetilde{H}_\alpha$ satisfy:
\begin{equation}
	\begin{split}
		M^\dagger Mv_2=E^2v_2,\;
		v_1=\frac{1}{E}Mv_2.
	\end{split}
\end{equation}
Notably, $v_1$ also satisfies $M M^\dagger v_1=E^2v_1$. Here,
\begin{equation}
	\begin{split}
		M^\dagger M&=|\vec{J}_1|^2+|\vec{J}_2|^2+t_1^2+t_2^2+t_3^2+t_4^2+2(t_1J_{1x}+\alpha t_2J_{1y}+t_3J_{2x}+\alpha t_4J_{2y})\widetilde{\tau_1}\\&+2(\alpha t_1J_{1y}-t_2J_{1x}+ \alpha t_3J_{2y}- t_4J_{2x})\widetilde{\tau_2}+2(-\vec{J}_1\cdot\vec{J}_2+t_1t_3+t_2t_4)\widetilde{\tau_3}\\
		MM^\dagger &=|\vec{J}_1|^2+|\vec{J}_2|^2+t_1^2+t_2^2+t_3^2+t_4^2+2(t_1J_{1x}-\alpha t_2J_{1y}-t_3J_{2x}+\alpha t_4J_{2y})\widetilde{\tau_1}\\&+2(\alpha t_1J_{1y}+t_2J_{1x}-\alpha t_3J_{2y}-t_4J_{2x})\widetilde{\tau_2}+2(+\vec{J}_1\cdot\vec{J}_2+t_1t_3+t_2t_4)\widetilde{\tau_3}.
	\end{split}
\end{equation}
One can then reproduce the dispersion by diagonalizing $M^\dagger M$:\begin{equation}
	\begin{split}
		E_{\alpha\beta\gamma}&=\epsilon_0+\gamma\bigl\{\vec{J}_1^2+\vec{J}_2^2+|{\bf t}|^2+2\beta\bigl[(t_1t_3+t_2t_4)^2+(\vec{J}_1^2+\vec{J}_2^2)|{\bf t}|^2/2\\&+(\vec{J}_1^2-\vec{J}_2^2)(t_1^2+t_2^2-t_3^2-t_4^2)/2+ (\vec{J}_1\cdot\vec{J}_2)^2+2\alpha (t_1t_4-t_2t_3)|\vec{J}_1\times\vec{J}_2|\bigr]^{1/2}\bigr\}^{1/2}, 
	\end{split}
\end{equation}
We now turn to the Berry curvature. Notably, $v_1$ and $v_2$ have the same norm:
\begin{equation}
	\begin{split}
		v_1^\dagger v_1=\frac{1}{E^2}v_2^\dagger M^\dagger Mv_2=\frac{1}{E^2}v_2^\dagger (E^2)v_2=v_2^\dagger v_2.
	\end{split}
\end{equation}
That says, after normalizing the eigenvector, $v_1^\dagger v_1=v_2^\dagger v_2=1/2$. The normalization of $v_1$ and $v_2$ are thus independent. The Berry connection (and Berry curvature) can be decomposed into contributions from $v_1$ and $v_2$:
\begin{equation}
	\begin{split}
		\vec{A}_i=i[v_1^\dagger ,v_2^\dagger]\partial_i [v_1,v_2]=iv_1^\dagger\partial_iv_1+iv_2^\dagger\partial_iv_2.
	\end{split}
\end{equation}
The contribution from $v_1$ (or $v_2$) can be determined from $M M^\dagger$ (or $M^\dagger M$). For a general $2\times 2$ Hamiltonian $H=h_1\tau_x+h_2\tau_y+h_3\tau_z$, the Berry curvature is $\Omega_{ij}=\frac{1}{2h^3}\sum_{a,b,c=1}^3\epsilon_{abc}h_a\partial_ih_b\partial_jh_c$. We can then apply this formula to obtain the contribution from $v_1$ and $v_2$, with an extra prefactor of $1/4$ from their normalization. For the nematic state  
$(\vec{J}_1,\vec{J}_2)=(\frac{{\bf S_1+S_2}}{2},\frac{{\bf S_1-S_2}}{2})=J(\hat{x},0)$ and odd-parity spin-splitting state $(\vec{J}_1,\vec{J}_2)=J(\hat{x},\hat{y})$, these two contribution exactly cancel each other, leading to zero Berry curvature. For the scalar odd-parity state 
$(\vec{J}_1,\vec{J}_2)=J(\hat{x},\hat{x})$, the Berry curvature is 
\begin{equation}
	\begin{split}
		\Omega^{ij}_{\alpha\beta\gamma}=\frac{\beta}{|\vec{h}|^3}\sum_{abc}\epsilon_{abc}[h_a\partial_{k_i}&h_b\partial_{k_j}h_c+\widetilde{h}_a\partial_{k_i}\widetilde{h}_b\partial_{k_j}\widetilde{h}_c]
		\\\vec{h}\equiv(Jt_x,Jt_y,J^2+(-t_x^2-t_y^2+\widetilde{t}_x^2+\widetilde{t}_y^2)/4),&\;\;\vec{\widetilde{h}}\equiv(J\widetilde{t}_x,J\widetilde{t}_y,J^2+(+t_x^2+t_y^2-\widetilde{t}_x^2-\widetilde{t}_y^2)/4)
	\end{split}
\end{equation}
The inversion symmetry breaking is on the order of $J^2$, depending on $t_xt_y-\widetilde{t_x}\widetilde{t_y}$. For example, in SG129(2c), $t_x=t_{x0}\cos\frac{k_x}{2}\cos\frac{k_y}{2}$ and $t_y=t_{y0}\cos\frac{k_x}{2}\cos\frac{k_y}{2}\sin k_z$. With ${\bf Q}=(\pi,0,0)$, $\widetilde{t_x}=-t_{x0}\sin\frac{k_x}{2}\cos\frac{k_y}{2}\sin k_z$ and $\widetilde{t_y}=-t_{u0}\sin\frac{k_x}{2}\cos\frac{k_y}{2}$. We obtain $(t_xt_y-\widetilde{t_x}\widetilde{t_y})\propto\sin k_z$, in agreement with the symmetry requirements from the phenomenological result in Table.\ref{T:3}. 

\section{Class 2}
In this section, we repeat the above analysis for the second inversion class, where the atoms are located at inversion center. The Hamiltonian of the doubled unit cell is:
\begin{equation}
	\begin{split}
		H&=\begin{pmatrix}
			h_0({\bf k})&O_M\\O_M^\dagger&\widetilde{h_0}({\bf k+Q})
		\end{pmatrix}=\begin{pmatrix}
			\epsilon_0+t_0+t_x\tau_x+t_z\tau_z & \vec{J}_1\cdot\vec{\sigma}+\tau_z\vec{J}_2\cdot\vec{\sigma}\\\vec{J}_1\cdot\vec{\sigma}+\tau_z\vec{J}_2\cdot\vec{\sigma}&\epsilon_0-t_0+\widetilde{t}_y\tau_y+\widetilde{t}_z\tau_z
		\end{pmatrix}\\
		&\equiv  \epsilon_0+t_1\rho_0\tau_x+t_2\rho_0\tau_y+t_3\rho_z\tau_x+t_4\rho_z\tau_y+t_5\rho_0\tau_z+t_6\rho_z\tau_z+t_0\rho_z+\rho_x(\vec{J}_1\cdot\vec{\sigma}+\tau_z\vec{J}_2\cdot\vec{\sigma})
	\end{split}
\end{equation}
where $t_1=t_3 \equiv t_x/2$, $t_2=-t_4 \equiv \widetilde{t_y}/2$, $t_5 \equiv (t_z+ \widetilde{t_z})/2$, and $t_6 \equiv (t_z- \widetilde{t_z})/2$. Pauli matrix $\tau_z=\pm$ distinguishes sublattices within each nonmagnetic unit cell, and $\rho_x=\pm$ differentiates nonmagnetic unit cells when unit cell doubling is applied. Time-reversal operator and inversion operator are $T=i\sigma_yK$ and $I=\frac{1+\rho_z}{2}\otimes\tau_0+\frac{1-\rho_z}{2}\otimes\tau_z$, with momentum flip. 

$t_x$ and $t_z$ are k-even hopping terms in the nonmagnetic unit cells, listed in Table.\ref{T:5}. $t_x$ is the nearest neighbored hopping, and $t_z$ hopping is generically weaker, since it comes from further neighbored hopping. For every k-even $t_x\tau_x$ terms in $h_0({\bf k})$, we obtain the k-odd $\widetilde{t_y}\tau_y$ term from $h_0({\bf k+Q})$, with the correct operators $\tau_{y}$ due to time-reversal symmetry. For every k-even $t_z\tau_z$ term in $h_0({\bf k})$, we obtain the k-even $\widetilde{t_z}\tau_z$ term from $h_0({\bf k+Q})$. In the following analysis, we will take $t_0=0$ and drop the trivial $\epsilon_0$ term for simplicity.

The three AFM states are described by $(\vec{J}_1,\vec{J}_2)=(\frac{{\bf S_1+S_2}}{2},\frac{{\bf S_1-S_2}}{2})=J(\hat{x},\hat{x}),J(\hat{x},0),\text{and }J(\hat{x},\hat{y})$.Without loss of generality, we will consider $\vec{J}_{1,2}$ in the xy plane. Using $[\rho_z\sigma_z,H]=0$, we can block-diagonalize the Hamiltonian:
\begin{equation}
	\begin{split}
		&u^\dagger Hu=\left(\begin{array}{cc}
			H_{\alpha=+}&  \\
			&H_{\alpha=-} 
		\end{array}\right),\;u=\left(\begin{array}{cccc}
			1&&&\\&&1&\\&&&1\\&1&&
		\end{array}\right),\\&H_\alpha=t_1\widetilde{\rho_0}\tau_x+t_2\widetilde{\rho_0}\tau_y+t_3\widetilde{\rho_z}\tau_x+t_4\widetilde{\rho_z}\tau_y+t_5\widetilde{\rho_0}\tau_z+t_6\widetilde{\rho_z}\tau_z+t_0\widetilde{\rho_z}+\vec{J}_{1\alpha}\cdot\vec{\widetilde{\rho}}+\tau_z\vec{J}_{2\alpha}\cdot\vec{\widetilde{\rho}},
	\end{split}
\end{equation}
where the unitary transformation $u$ is in $(\rho,\sigma)$ space, and $\vec{J}_{i\alpha}\equiv(J_{ix},\alpha J_{iy},0)$.

For the first two AFM orders, the band remains doubly degenerate. For the ${\bf S_1} \times {\bf S_2} \neq 0$ state, the dispersion is:

\begin{equation}
	\begin{split}
		E_{\alpha\beta\gamma}=\gamma\left[2J^2+|{\bf t}|^2+2\beta\sqrt{J^2(|{\bf t}|^2+t_5^2-t_6^2)+2\alpha J^2(t_1t_4-t_2t_3)+(t_1t_3+t_2t_4+t_5t_6)^2}\right]^{1/2}, 
	\end{split}
\end{equation}
where $\alpha,\beta,\gamma=\pm$ and $|{\bf t}|^2\equiv t_1^2+t_2^2+t_3^2+t_4^2+t_5^2+t_6^2$. The magnitude of spin splitting depends on $J\sqrt{t_1t_4-t_2t_3}\propto J(t_x\widetilde{t_y})^{1/2}$.

We now turn to the Berry curvature dipole of the collinear scalar odd-parity state $(\vec{J}_1,\vec{J}_2)=(\frac{{\bf S_1+S_2}}{2},\frac{{\bf S_1-S_2}}{2})=J(\hat{x},0)$, the dispersion is 
\begin{equation}
	\begin{split}
		E_{\beta\gamma}=\gamma\left[J^2+|{\bf t}|^2+2\beta\sqrt{J^2(t_1^2+t_2^2+t_5^2)+(t_1t_3+t_2t_4+t_5t_6)^2}\right]^{1/2},
	\end{split}
\end{equation}

The general formula for Berry curvature is 
$\Omega_{ij}=-2\Im{\text{Tr}[(\partial_iP_n)(1-P_n)(\partial_jP_n)]}$, where $P_n$ is the projection operator onto band n. 
Here, for each $4\times4$ Hamiltonian $H_\alpha$, the projection operator is
\begin{equation}
	\begin{split}
		&P_{\alpha\beta\gamma}=\frac{1}{4}\left[1+\frac{H_\alpha}{E_{\beta\gamma}}\right]\left[1+\frac{\bar{H}_\alpha}{\bar{E}_{\beta}}\right],\\
		&\bar{H}_\alpha\equiv2\left[(t_1t_3+t_2t_4+t_5t_6)\widetilde{\rho_z}\tau_0+Jt_1\widetilde{\rho_x}\tau_x+Jt_2\widetilde{\rho_x}\tau_y+Jt_5\widetilde{\rho_x}\tau_z\right], \\&\bar{E}_{\beta}\equiv2\beta\sqrt{J^2(t_1^2+t_2^2+t_5^2)+(t_1t_3+t_2t_4+t_5t_6)^2}
	\end{split}
\end{equation}
The Berry curvature is
\begin{equation}
	\begin{split}
		&\bar{\beta}=+:\Omega^{ij}_{\alpha\beta\gamma}=\frac{J^2}{16\bar{E}_\beta^3E^3_{\alpha\beta\gamma}}[\partial_it_z(3t_x^5\partial_j\widetilde{t}_y-9t_x^4\widetilde{t}_y\partial_jt_x+6t_x^3\widetilde{t}_y^2\partial_j\widetilde{t}_y+2t_x^2\widetilde{t}_y^3\partial_jt_x-t_x\widetilde{t}_y^4\partial_j\widetilde{t}_y-\widetilde{t}_y^5\partial_jt_x)\\&-t_z\partial_it_x\partial_j\widetilde{t}_y(3t_x^4+6t_x^2\widetilde{t}_y^2-\widetilde{t}_y^4)-8\widetilde{t_z}t_x^4\partial_it_x\partial_j\widetilde{t}_y+\partial_i\widetilde{t_z}(4t_x^5\partial_j\widetilde{t}_y-8t_x^4\widetilde{t}_y\partial_jt_x+4t_x^3\widetilde{t}_y^2\partial_j\widetilde{t}_y)-(i\leftrightarrow j)],\\&\bar{\beta}=-:\Omega^{ij}_{\alpha\beta\gamma}=\frac{J^2}{16\bar{E}_\beta^3E^3_{\alpha\beta\gamma}}[\partial_i\widetilde{t_z}(-3\widetilde{t}_y^5\partial_jt_x+9t_x\widetilde{t}_y^4\partial_j\widetilde{t}_y-6t_x^2\widetilde{t}_y^3\partial_jt_x-2t_x^3\widetilde{t}_y^2\partial_j\widetilde{t}_y+t_x^4\widetilde{t}_y\partial_jt_x+t_x^5\partial_j\widetilde{t}_y)\\&-\widetilde{t_z}\partial_it_x\partial_j\widetilde{t}_y(3\widetilde{t}_y^4+6t_x^2\widetilde{t}_y^2-t_x^4)-8t_z\widetilde{t}_y^4\partial_it_x\partial_j\widetilde{t}_y+\partial_it_z(-4\partial_jt_x\widetilde{t}_y^5+8\partial_j\widetilde{t}_yt_x\widetilde{t}_y^4-4\partial_jt_xt_x^2\widetilde{t}_y^3)-(i\leftrightarrow j)]
	\end{split}
\end{equation}
where $\bar{\beta}\equiv\beta\text{sgn}(t_x^2-\widetilde{t}_y^2)$. We have kept the leading contributions, quadratic in $J$ and linear in the further hopping strength $t_z$ (or $\widetilde{t_z}$). The symmetry of Berry curvature follows $J^2t_x\widetilde{t_y}t_z$ and $J^2t_x\widetilde{t_y}\widetilde{t_z}$. Such k-odd Berry curvature allows non-zero Berry curvature dipoles $\langle d\Omega^{ij}/dk_m\rangle_{FS}$. 

\section{List of materials}
In this section, we list the materials where the Landau theory is applicable. The symbol $*$ denotes materials where certain magnetic atoms can be described by the microscopic models. Space groups, wavevectors, and Magndata entry labels are included. The category Scalar and/or spin odd-parity order appears because SOC makes the precise categorization of the non-relativistic state uncertain and both possibilities are consistent with the experimental structure.

\begin{table}[h]
	\begin{tabular}{|c|c|}
		\hline
		\multirow{5}{*}{Spin odd-parity} & DyMn$_2$O$_5$(55X,1.324), Tm$_5$Pt$_2$In$_4$(55T,1.790*), CoNb$_2$O$_6$(60Y,1.224), \\ \cline{2-2} 
		& La$_{3/8}$Ca$_{5/8}$MnO$_3$(62Z,1.173), La$_{1/3}$Ca$_{2/3}$MnO$_3$(62Z,1.174), \\ \cline{2-2} 
		& La$_{1/3}$Ca$_{2/3}$MnO$_3$(62X,1.175), Gd$_2$BaCuO$_5$(62Z,1.443), Er$_2$Pt(62X,1.444), \\ \cline{2-2} 
		& CeNiAsO(129X,1.272*), Sr$_2$FeO$_3$Cl(129M,1.380,1.382*), Sr$_2$FeO$_3$Br(129M,1.381,1.383*), \\ \cline{2-2} 
		& Sr$_2$FeO$_3$F(129M,1.385,1.386,1.387*), NiCr$_2$O$_4$(141M,1.688), Dy$_2$Co$_3$Al$_9$(63Z,1.267) \\ \hline
		\multirow{7}{*}{Scalar odd-parity} & Na$_2$MnF$_5$(14Z,1.55*),  Lu$_2$MnCoO$_6$(14A,1.32*), YbLuCoMnO$_6$(14B,1.329*), \\ \cline{2-2} 
		& Lu$_2$CoMnO$_6$(14B,1.330*), Yb$_2$CoMnO$_6$(14B,1.328*), SmMn$_2$O$_5$(55X,1.192), \\ \cline{2-2} 
		& DyMn$_2$O$_5$ (55X,1.599,1.76), PrMn$_2$O$_5$(55X,1.325), Sr$_2$Fe$_3$Se$_2$O$_3$(55U,1.463,1.626), \\ \cline{2-2} 
		& Tb$_5$Pd$_2$In$_4$(55Y,1.697*), FeNb$_2$O$_6$(60Y,1.655), BaCdVO(PO$_4$)$_2$(61Y,1.298), \\ \cline{2-2} 
		& LuMnO$_3$(62X,1.101), HoMnO$_3$(62X,1.20), PrNiO$_3$(62T,1.43), NdNiO$_3$(62T,1.45), \\ \cline{2-2} 
		& TmMnO$_3$(62X,1.341), SmNiO$_3$(62T,1.353), EuNiO$_3$(62T,1.354), HoNiGe(62T,1.374), \\ \cline{2-2} 
		& Na$_2$CuSO$_4$Cl$_2$(62X,1.682), NiCr$_2$O$_4$(141M,1.685), Na$_2$Ni$_2$TeO$_6$(193L,1.646*), MnS$_2$(205X,1.18) \\ \hline
		\multirow{3}{*}{Scalar and/or  spin odd-parity} & PrMn$_2$O$_5$(55X,1.19), GdMn$_2$O$_5$(55X,1.54,1.299,1.300), BiMn$_2$O$_5$(55U,1.74,1.75), \\ \cline{2-2} 
		& Tm$_5$Ni$_2$In$_4$(55T,1.170*), NdNiO$_3$(62T,1.44), Cs$_2$CoCl$_4$(62T,1.51),LuMnO$_3$(62X,1.340) \\ \hline
		{Nematic} &  Fe(ND$_3$)$_2$PO$_4$ (14B,1.66),  NaMnGe$_2$O$_6$(15A,1.260), PrFe$_2$Al$_8$(55U,1.681), VOCl(59R,1.37), \\ \cline{2-2} 
		& MnV$_2$O$_6$(60Z,1.196), NiNb$_2$O$_6$(60Y,1.654), CoNb$_2$O$_6$(60Z,1.656), BaNd$_2$O$_4$(62T,1.95), \\ \cline{2-2}
		& ErNiGe(62Y,1.379), Y$_2$BaCuO$_5$(62T,1.445), SrNd$_2$O$_4$(62T,1.577), DyBaCuO$_5$(62Y,1.650), \\ \cline{2-2}& 
		Dy$_2$TiO$_5$(62Y,1.698), PrPdSn(62T,1.744)
		, Ho$_2$BaCuO$_5$(62Y,1.651), DyGe(63Y,1.361)  \\ \cline{2-2}& Li$_2$VOSiO$_4$(129M,1.9), Sr$_2$CoO$_3$Cl(129M,1.389), Fe$_{1.05}$Te(129R,1.434), \\ \cline{2-2}&  NdNiMg$_{15}$ (129M,1.457), HoSbTe(129X,1.749), Dy$_3$Ru$_4$Al$_12$ (194L,1.115)\\ \hline \end{tabular}
\end{table}



\begin{table*}[htb]
	\caption{Little co-group, momentum-dependent spin-splitting, space group and wavevectors. Labels $^*$ denote spin splittings not along the conventional axis. For instance, $11C_1$ has spin splitting along $k_y$ instead of $k_z$.  }
	\label{tab:spacegroups}
	\begin{tabular}{c|c|c}
		\toprule
		$P$ & Spin-splitting &Space group \& Wavevector\\ \hline
		
		$C_{2h}$& $k_z$ &$11(C^{*}_{1},D^{*}_{1},E^{*}_{1},Z^{*}_{1}),14(C^{*}_{1},Z^{*}_{1}),63(R^{}_{1}),176(L^{}_{1})$\\ \hline
		
		$C_{4h}$&$k_z$ &$84(A^{}_{1},Z^{}_{1}),85(A^{}_{1},M^{}_{1}),86(M^{}_{1},Z^{}_{1}),88(M^{}_{1})$\\ \hline
		
		$C_{6h}$& $k_z$ &$176(A^{}_{1})$\\ \hline
		
		$D_{2h}$& $k_{z,y,x}$ & $\begin{matrix}
			48(S^{}_{1,2},T^{}_{1,2},U^{}_{1,2},X^{}_{1,2},Y^{}_{1,2},Z^{}_{1,2}),49(R^{}_{1,2},T^{}_{1,2},U^{}_{1,2},Z^{}_{1,2}),50(R^{}_{1,2},S^{}_{1,2},T^{}_{1,2},U^{}_{1,2},X^{}_{1,2},Y^{}_{1,2}),\\51(R^{}_{1,2},S^{}_{1,2},U^{}_{1,2},X^{}_{1,2}),52(R^{}_{1,2},U^{}_{1,2},X^{}_{1,2},Y^{}_{1,2},Z^{}_{1,2}),53(S^{}_{1,2},T^{}_{1,2},X^{}_{1,2},Z^{}_{1,2}),\\54(S^{}_{1,2},T^{}_{1,2},X^{}_{1,2},Z^{}_{1,2}),55(T^{}_{1,2},U^{}_{1,2},X^{}_{1,2},Y^{}_{1,2}),56(X^{}_{1,2},Y^{}_{1,2},Z^{}_{1,2}),57(S^{}_{1,2},U^{}_{1,2},Y^{}_{1,2},Z^{}_{1,2}),\\58(X^{}_{1,2},Y^{}_{1,2},Z^{}_{1,2}),59(T^{}_{1,2},U^{}_{1,2},X^{}_{1,2},Y^{}_{1,2}),60(X^{}_{1,2},Y^{}_{1,2},Z^{}_{1,2}),61(X^{}_{1,2},Y^{}_{1,2},Z^{}_{1,2}),\\62(X^{}_{1,2},Y^{}_{1,2},Z^{}_{1,2}),63(T^{}_{1,2},Z^{}_{1,2}),64(T^{}_{1,2},Z^{}_{1,2}),66(T^{}_{1,2},Z^{}_{1,2}),68(T^{}_{1,2},Z^{}_{1,2}),70(T^{}_{1,2},Y^{}_{1,2},Z^{}_{1,2}),\\124(R^{}_{1,2}),125(R^{}_{1,2},X^{}_{1,2}),126(R^{}_{1,2},X^{}_{1,2}),127(R^{}_{1,2},X^{}_{1,2}),128(X^{}_{1,2}),129(R^{}_{1,2},X^{}_{1,2}),130(X^{}_{1,2}),\\132(R^{}_{1,2}),133(R^{}_{1,2},X^{}_{1,2}),134(R^{}_{1,2},X^{}_{1,2}),135(R^{}_{1,2},X^{}_{1,2}),136(X^{}_{1,2}),137(R^{}_{1,2},X^{}_{1,2}),138(X^{}_{1,2}),\\141(X^{}_{1,2}),142(X^{}_{1,2}),192(L^{}_{1,2}),193(L^{}_{1,2}),194(L^{}_{1,2}),201(M^{}_{1,2},X^{}_{1,2}),203(X^{}_{1,2}),205(X^{}_{1,2}),230(N^{}_{1,2})\end{matrix}$ \\ \hline

		$D_{4h}$ & $k_z$ & $\begin{matrix}
			124(A^{}_{1,2},Z^{}_{1,2}),125(A^{}_{3,4},M^{}_{3,4}),126(A^{}_{3,4},M^{}_{3,4},Z^{}_{1,2}),128(Z^{}_{1,2}),130(Z^{}_{1,2}),131(A^{}_{1,2},Z^{}_{1,2}),\\132(A^{}_{1,2},Z^{}_{1,2}),133(A^{}_{1,2},M^{}_{1,2},Z^{}_{1,2}),134(M^{}_{1,2},Z^{}_{1,2}),135(Z^{}_{1,2}),136(Z^{}_{1,2}),137(Z^{}_{1,2}),138(Z^{}_{1,2}),\\141(M^{}_{1,2}),142(M^{}_{1,2}),222(M^{}_{3,4},X^{*}_{1,2}),223(X^{*}_{1,2}),224(M^{}_{1,2},X^{*}_{1,2}),227(X^{*}_{1,2}),228(X^{*}_{1,2})
		\end{matrix}$\\ \hline
		
		$D_{3d}$ &$k_z$ & $163(A^{}_{3}),165(A^{}_{3}),167(T^{}_{3}),226(L^{*}_{3}),228(L^{*}_{3})$  \\ \hline
		
		$D_{6h}$ & $k_z$ & $192(A^{}_{5,6}),193(A^{}_{1,2}),194(A^{}_{1,2})$  \\ \hline
		
		$C_{2h}$ & $c_1k_x+c_2k_y$& $\begin{matrix}13(A^{*}_{1},B^{*}_{1},D^{*}_{1},E^{*}_{1}),14(A^{*}_{1},B^{*}_{1}),15(A^{*}_{1},M^{*}_{1}),64(S^{}_{1}),67(R^{}_{1},S^{}_{1}),68(R^{}_{1},S^{}_{1}),72(S^{*}_{1},R^{*}_{1}),73(T^{}_{1},S^{*}_{1},R^{*}_{1}),\\74(T^{}_{1}),85(R^{}_{1},X^{}_{1}),86(R^{}_{1},X^{}_{1}),88(X^{}_{1}),140(N^{*}_{1}),142(N^{*}_{1}),163(L^{*}_{1}),165(L^{*}_{1}),167(L^{*}_{1}),206(N^{}_{1})
		\end{matrix}$
		\\ \hline
		
		$C_{4h}$&  $\begin{matrix}
			c_1k_xk_yk_z+\\c_2k_z(k_x^2-k_y^2)
		\end{matrix}$ &$84(A^{}_{2},Z^{}_{2}),85(A^{}_{2},M^{}_{2}),86(M^{}_{2},Z^{}_{2}),88(M^{}_{2})$\\ \hline
		
		$D_{4h}$ & $k_z(k_x^2-k_y^2)$ & $\begin{matrix}125(A^{}_{1,2},M^{}_{1,2}),126(M^{}_{1,2}),132(A^{}_{3,4},Z^{}_{3,4}),133(M^{}_{3,4}),134(M^{}_{3,4},Z^{}_{3,4}),136(Z^{}_{3,4}),\\138(Z^{}_{3,4}),222(M^{}_{1,2}),224(M^{}_{3,4},X^{*}_{3,4}),227(X^{*}_{3,4}),228(X^{*}_{3,4})\end{matrix}$\\ \hline
		
		$D_{2h}$ & $k_xk_yk_z$ & $56(S^{}_{1,2}),58(R^{}_{1,2}),59(R^{}_{1,2},S^{}_{1,2}),62(T^{}_{1,2})$  \\ \hline

		$D_{4h}$ & $k_xk_yk_z$ & $\begin{matrix}126(A^{}_{1,2}),128(A^{}_{1,2}),129(A^{}_{3,4},M^{}_{3,4}),130(M^{}_{3,4}),131(A^{}_{3,4},Z^{}_{3,4}),133(Z^{}_{3,4}),135(Z^{}_{3,4}),\\136(A^{}_{1,2}),137(A^{}_{1,2},M^{}_{3,4},Z^{}_{3,4}),138(M^{}_{3,4}),141(M^{}_{3,4}),142(M^{}_{3,4}),223(X^{*}_{3,4})
		\end{matrix}$\\ \hline
		
		$O_h$ & $k_xk_yk_z$ & $222(R^{}_{1}),223(R^{}_{1}),230(H^{}_{1})$  \\ \hline
		
		$D_{4h}$ & $\begin{matrix}
			k_xk_yk_z\times\\(k_x^2-k_y^2)
		\end{matrix}$ & $129(A^{}_{1,2},M^{}_{1,2}),130(M^{}_{1,2}),136(A^{}_{3,4}),137(M^{}_{1,2}),138(M^{}_{1,2})$ \\ \hline
		\bottomrule
	\end{tabular}
	\label{T:1}
\end{table*}

\begin{table*}[htb]
	\caption{Little co-group, even-parity order, space group and wavevectors. Labels $^*$ denote orders not along the conventional axis. For instance, 11$C_1$ hosts nematicity $c_1k_xk_y+c_2k_zk_y$.  }
	\begin{tabular}{c|c|c|c}
		\toprule
		$P$  &IR of P& Nematicity&Space group \& Wavevector\\ \hline
		
		$C_{2h}$& $B_g$&$\begin{matrix}c_1k_xk_z\\+c_2k_yk_z\end{matrix}$ &$\begin{matrix}11(Z^{*}_{1},C^{*}_{1},D^{*}_{1},E^{*}_{1}),13(B^{*}_{1},D^{*}_{1},A^{*}_{1},E^{*}_{1}),14(Z^{*}_{1},B^{*}_{1},C^{*}_{1},A^{*}_{1}),15(A^{*}_{1},M^{*}_{1}),63(R^{}_{1}),\\64(S^{}_{1}),67(S^{}_{1},R^{}_{1}),68(S^{}_{1},R^{}_{1}),72(S^{*}_{1},R^{*}_{1}),73(S^{*}_{1},R^{*}_{1},T^{}_{1}),74(T^{}_{1}),85(X^{}_{1},R^{}_{1}),\\86(X^{}_{1},R^{}_{1}),88(X^{}_{1}),140(N^{*}_{1}),142(N^{*}_{1}),163(L^{*}_{1}),165(L^{*}_{1}),167(L^{*}_{1}),176(L^{}_{1}),206(N^{}_{1})\end{matrix}$\\ \hline
		
		$C_{4h}$& $B_g$&$\begin{matrix}c_1(k_x^2-k_y^2)\\+c_2k_xk_y\end{matrix}$&$84(Z^{}_{1,2},A^{}_{1,2}),85(M^{}_{1,2},A^{}_{1,2}),86(Z^{}_{1,2},M^{}_{1,2}),88(M^{}_{1,2})$\\ \hline
		
		$C_{6h}$& $B_g$&$k_xk_z(k_x^2-3k_y^2)$&$176(A^{}_{1})$\\ \hline
		
		$D_{2h}$& $\begin{matrix}B_{1g}\\B_{2g}\\B_{3g}\end{matrix}$&$\begin{matrix}k_xk_y\\k_zk_x\\k_yk_z\end{matrix}$&$\begin{matrix}48(X^{}_{1,2},Z^{}_{1,2},U^{}_{1,2},Y^{}_{1,2},S^{}_{1,2},T^{}_{1,2}),49(Z^{}_{1,2},U^{}_{1,2},T^{}_{1,2},R^{}_{1,2}),\\50(X^{}_{1,2},U^{}_{1,2},Y^{}_{1,2},S^{}_{1,2},T^{}_{1,2},R^{}_{1,2}),51(X^{}_{1,2},U^{}_{1,2},S^{}_{1,2},R^{}_{1,2}),52(X^{}_{1,2},Z^{}_{1,2},U^{}_{1,2},Y^{}_{1,2},R^{}_{1,2}),\\53(X^{}_{1,2},Z^{}_{1,2},S^{}_{1,2},T^{}_{1,2}),54(X^{}_{1,2},Z^{}_{1,2},S^{}_{1,2},T^{}_{1,2}),55(X^{}_{1,2},U^{}_{1,2},Y^{}_{1,2},T^{}_{1,2}),\\56(X^{}_{1,2},Z^{}_{1,2},Y^{}_{1,2},S^{}_{1,2}),57(Z^{}_{1,2},U^{}_{1,2},Y^{}_{1,2},S^{}_{1,2}),58(X^{}_{1,2},Z^{}_{1,2},Y^{}_{1,2},R^{}_{1,2}),\\59(X^{}_{1,2},U^{}_{1,2},Y^{}_{1,2},S^{}_{1,2},T^{}_{1,2},R^{}_{1,2}),60(X^{}_{1,2},Z^{}_{1,2},Y^{}_{1,2}),61(X^{}_{1,2},Z^{}_{1,2},Y^{}_{1,2}),\\62(X^{}_{1,2},Z^{}_{1,2},Y^{}_{1,2},T^{}_{1,2}),63(T^{}_{1,2},Z^{}_{1,2}),64(T^{}_{1,2},Z^{}_{1,2}),66(T^{}_{1,2},Z^{}_{1,2}),68(T^{}_{1,2},Z^{}_{1,2}),\\70(T^{}_{1,2},Z^{}_{1,2},Y^{}_{1,2}),124(R^{}_{1,2}),125(R^{}_{1,2},X^{}_{1,2}),126(R^{}_{1,2},X^{}_{1,2}),127(R^{}_{1,2},X^{}_{1,2}),128(X^{}_{1,2}),\\129(R^{}_{1,2},X^{}_{1,2}),130(X^{}_{1,2}),132(R^{}_{1,2}),133(R^{}_{1,2},X^{}_{1,2}),134(R^{}_{1,2},X^{}_{1,2}),135(R^{}_{1,2},X^{}_{1,2}),\\136(X^{}_{1,2}),137(R^{}_{1,2},X^{}_{1,2}),138(X^{}_{1,2}),141(X^{*}_{1,2}),142(X^{*}_{1,2}),\\192(L^{*}_{1,2}),193(L^{*}_{1,2}),194(L^{*}_{1,2}),201(M^{}_{1,2},X^{}_{1,2}),203(X^{}_{1,2}),205(X^{}_{1,2}),230(N^{*}_{1,2})\end{matrix}$\\ \hline
		
		$D_{3d}$& $A_{2g}$&$k_xk_z(k_x^2-3k_y^2)$&$163(A^{}_{3}),165(A^{}_{3}),167(T^{}_{3}),226(L^{*}_{3}),228(L^{*}_{3})$\\ \hline
		
		$D_{4h}$& $B_{1g}$&$k_x^2-k_y^2$&$\begin{matrix}126(A^{}_{1,2,3,4}),131(Z^{}_{1,2,3,4},A^{}_{1,2,3,4}),133(Z^{}_{1,2,3,4}),135(Z^{}_{1,2,3,4}),\\137(Z^{}_{1,2,3,4}),141(M^{}_{1,2,3,4}),142(M^{}_{1,2,3,4}),223(X^{*}_{1,2,3,4})\end{matrix}$\\ \hline
		
		$D_{4h}$& $B_{2g}$&$k_xk_y$&$\begin{matrix}125(M^{}_{1,2,3,4},A^{}_{1,2,3,4}),126(M^{}_{1,2,3,4}),129(M^{}_{1,2,3,4},A^{}_{1,2,3,4}),130(M^{}_{1,2,3,4}),\\132(Z^{}_{1,2,3,4},A^{}_{1,2,3,4}),133(M^{}_{1,2,3,4}),134(Z^{}_{1,2,3,4},M^{}_{1,2,3,4}),136(Z^{}_{1,2,3,4},A^{}_{1,2,3,4}),\\137(M^{}_{1,2,3,4}),138(Z^{}_{1,2,3,4},M^{}_{1,2,3,4}),222(M^{}_{1,2,3,4}),224(M^{}_{1,2,3,4},X^{*}_{1,2,3,4}),227(X^{*}_{1,2,3,4}),228(X^{*}_{1,2,3,4})\end{matrix}$\\ 
		\hline
		
		$D_{4h}$& $A_{2g}$&$k_xk_y(k_x^2-k_y^2)$&$124(Z^{}_{1,2},A^{}_{1,2}),126(Z^{}_{1,2}),128(Z^{}_{1,2},A^{}_{1,2}),130(Z^{}_{1,2}),133(A^{}_{1,2}),137(A^{}_{1,2}),222(X^{*}_{1,2})$\\ \hline
		
		$D_{6h}$& $A_{2g}$&$\begin{matrix}k_xk_y(k_x^2-3k_y^2)\\ \times(k_y^2-3k_x^2)\end{matrix}$&$192(A^{}_{5,6})$\\ \hline
		
		$D_{6h}$& $B_{1g}$&$k_yk_z(k_y^2-3k_x^2)$&$193(A^{}_{1,2})$\\ \hline
		
		$D_{6h}$& $B_{2g}$&$k_xk_z(k_x^2-3k_y^2)$&$194(A^{}_{1,2})$\\ \hline
		
		$O_{h}$& $A_{2g}$&$\begin{matrix}(k_x^2-k_y^2)(k_y^2-k_z^2)\\ \times(k_z^2-k_x^2)\end{matrix}$&$222(R^{}_{1}),223(R^{}_{1}),230(H^{}_{1})$\\ \hline
		\bottomrule
	\end{tabular}
	\label{T:2}
\end{table*}

\begin{table*}[htb]
	\caption{Little co-group, spin-scalar odd-parity state, space group and wavevectors. }
	\begin{tabular}{c|c|c|c}
		\toprule
		$P$ & IR of $P$ & Symmetry &Space group \& Wavevector\\ \hline
		
		$C_{2h}$& $A_u$& $z$ &$\begin{matrix}13(A^{*}_{1},B^{*}_{1},D^{*}_{1},E^{*}_{1}),14(A^{*}_{1},B^{*}_{1}),15(A^{*}_{1},M^{*}_{1}),64(S^{}_{1}),67(R^{}_{1},S^{}_{1}),68(R^{}_{1},S^{}_{1}),\\72(S^{*}_{1},R^{*}_{1}),73(T^{}_{1},S^{*}_{1},R^{*}_{1}),74(T^{}_{1}),85(R^{}_{1},X^{}_{1}),86(R^{}_{1},X^{}_{1}),88(X^{}_{1}),140(N^{*}_{1}),\\142(N^{*}_{1}),163(L^{*}_{1}),165(L^{*}_{1}),167(L^{*}_{1}),206(N^{}_{1})\end{matrix}$\\ \hline
		
		$C_{4h}$& $A_u$& $z$ &$84(A^{}_{2},Z^{}_{2}),85(A^{}_{2},M^{}_{2}),86(M^{}_{2},Z^{}_{2}),88(M^{}_{2})$\\ \hline
		
		$D_{2h}$ & $\begin{matrix}
			B_{1u}\\B_{2u}\\B_{3u}\end{matrix}$ & $z,y,x$ & $\begin{matrix}
			51(R^{}_{1,2},S^{}_{1,2},U^{}_{1,2},X^{}_{1,2}),52(R^{}_{1,2},Y^{}_{1,2}),53(T^{}_{1,2},Z^{}_{1,2}),54(S^{}_{1,2},X^{}_{1,2}),\\55(T^{}_{1,2},U^{}_{1,2},X^{}_{1,2},Y^{}_{1,2}),56(S^{}_{1,2},X^{}_{1,2},Y^{}_{1,2}),57(S^{}_{1,2},U^{}_{1,2},Y^{}_{1,2},Z^{}_{1,2}),58(R^{}_{1,2},X^{}_{1,2},Y^{}_{1,2}),\\59(R^{}_{1,2},S^{}_{1,2},T^{}_{1,2},U^{}_{1,2},X^{}_{1,2},Y^{}_{1,2}),60(X^{}_{1,2},Z^{}_{1,2}),61(X^{}_{1,2},Y^{}_{1,2},Z^{}_{1,2}),\\62(T^{}_{1,2},X^{}_{1,2},Y^{}_{1,2},Z^{}_{1,2}),63(T^{}_{1,2},Z^{}_{1,2}),64(T^{}_{1,2},Z^{}_{1,2}),127(R^{}_{1,2},X^{}_{1,2}),128(X^{}_{1,2}),\\129(R^{}_{1,2},X^{}_{1,2}),130(X^{}_{1,2}),135(R^{}_{1,2},X^{}_{1,2}),136(X^{}_{1,2}),137(R^{}_{1,2},X^{}_{1,2}),\\138(X^{}_{1,2}),193(L^{}_{1,2}),194(L^{}_{1,2}),205(X^{}_{1,2})
		\end{matrix}$ \\ \hline

		$D_{4h}$ & $A_{2u}$ & $z$ & $\begin{matrix}
			129(A^{}_{3,4},M^{}_{3,4}),130(M^{}_{3,4}),136(A^{}_{1,2}),137(M^{}_{3,4}),138(M^{}_{3,4})
		\end{matrix}$\\ \hline
		
		$C_{2h}$ & $B_{u}$ & $c_1x+c_2y$& $11(C^{*}_{1},D^{*}_{1},E^{*}_{1},Z^{*}_{1}),14(C^{*}_{1},Z^{*}_{1}),63(R^{}_{1}),176(L^{}_{1})$
		\\ \hline
		
		$C_{4h}$& $B_u$& $\begin{matrix}
			c_1xyz+\\c_2z(x^2-y^2)
		\end{matrix}$ &$84(A^{}_{1},Z^{}_{1}),85(A^{}_{1},M^{}_{1}),86(M^{}_{1},Z^{}_{1}),88(M^{}_{1})$\\ \hline
		
		$D_{4h}$ & $B_{2u}$ & $z(x^2-y^2)$ & $\begin{matrix}126(A^{}_{3,4}),128(A^{}_{1,2}),129(A^{}_{1,2},M^{}_{1,2}),130(M^{}_{1,2}),131(A^{}_{1,2},Z^{}_{1,2}),133(Z^{}_{1,2}),135(Z^{}_{1,2}),\\136(A^{}_{3,4}),137(A^{}_{1,2},M^{}_{1,2},Z^{}_{1,2}),138(M^{}_{1,2}),141(M^{}_{1,2}),142(M^{}_{1,2}),223(X^{*}_{1,2})\end{matrix}$\\ \hline
		
		$D_{2h}$ & $A_u$ & $xyz$ & $\begin{matrix}48(S^{}_{1,2},T^{}_{1,2},U^{}_{1,2},X^{}_{1,2},Y^{}_{1,2},Z^{}_{1,2}),49(R^{}_{1,2},T^{}_{1,2},U^{}_{1,2},Z^{}_{1,2}),\\50(R^{}_{1,2},S^{}_{1,2},T^{}_{1,2},U^{}_{1,2},X^{}_{1,2},Y^{}_{1,2}),52(U^{}_{1,2},X^{}_{1,2},Z^{}_{1,2}),53(S^{}_{1,2},X^{}_{1,2}),54(T^{}_{1,2},Z^{}_{1,2}),\\56(Z^{}_{1,2}),58(Z^{}_{1,2}),60(Y^{}_{1,2}),66(T^{}_{1,2},Z^{}_{1,2}),68(T^{}_{1,2},Z^{}_{1,2}),70(T^{}_{1,2},Y^{}_{1,2},Z^{}_{1,2}),124(R^{}_{1,2}),\\125(R^{}_{1,2},X^{}_{1,2}),126(R^{}_{1,2},X^{}_{1,2}),132(R^{}_{1,2}),133(R^{}_{1,2},X^{}_{1,2}),134(R^{}_{1,2},X^{}_{1,2}),\\141(X^{*}_{1,2}),142(X^{*}_{1,2}),192(L^{}_{1,2}),201(M^{}_{1,2},X^{}_{1,2}),203(X^{}_{1,2}),230(N^{*}_{1,2})\end{matrix}$  \\ \hline

		$D_{4h}$ & $B_{1u}$ & $xyz$ & $\begin{matrix}125(A^{}_{3,4},M^{}_{3,4}),126(M^{}_{3,4}),132(A^{}_{1,2},Z^{}_{1,2}),133(M^{}_{1,2}),134(M^{}_{1,2},Z^{}_{1,2}),\\136(Z^{}_{1,2}),138(Z^{}_{1,2}),222(M^{}_{3,4}),224(M^{}_{1,2},X^{*}_{1,2}),227(X^{*}_{1,2}),228(X^{*}_{1,2})
		\end{matrix}$\\ \hline
		
		$D_{4h}$ & $A_{1u}$ & $xyz(x^2-y^2)$ & $\begin{matrix}124(A^{}_{1,2},Z^{}_{1,2}),125(A^{}_{1,2},M^{}_{1,2}),126(A^{}_{1,2},M^{}_{1,2},Z^{}_{1,2}),128(Z^{}_{1,2}),130(Z^{}_{1,2}),\\131(A^{}_{3,4},Z^{}_{3,4}),132(A^{}_{3,4},Z^{}_{3,4}),133(A^{}_{1,2},M^{}_{3,4},Z^{}_{3,4}),134(M^{}_{3,4},Z^{}_{3,4}),135(Z^{}_{3,4}),\\136(Z^{}_{3,4}),137(Z^{}_{3,4}),138(Z^{}_{3,4}),141(M^{}_{3,4}),142(M^{}_{3,4}),222(M^{}_{1,2},X^{*}_{1,2}),\\223(X^{*}_{3,4}),224(M^{}_{3,4},X^{*}_{3,4}),227(X^{*}_{3,4}),228(X^{*}_{3,4})\end{matrix}$ \\ \hline
		
		$C_{6h}$& $B_u$&$\begin{matrix}c_1x(x^2-3y^2)\\+c_2y(y^2-3x^2)\end{matrix}$&$176(A^{}_{1})$\\ \hline
		
		$D_{3d}$& $A_{1u}$&$x(x^2-3y^2)$&$163(A^{}_{3}),165(A^{}_{3}),167(T^{}_{3}),226(L^{*}_{3}),228(L^{*}_{3})$\\ \hline
		
		$D_{6h}$& $A_{1u}$&$\begin{matrix}xyz(x^2-3y^2)\\ \times(y^2-3x^2)\end{matrix}$&$192(A^{}_{5,6})$\\ \hline
		
		$D_{6h}$& $B_{1u}$&$x(x^2-3y^2)$&$194(A^{}_{1,2})$\\ \hline
		
		$D_{6h}$& $B_{2u}$&$y(y^2-3x^2)$&$193(A^{}_{1,2})$\\ \hline
		
		$O_{h}$& $A_{1u}$&$\begin{matrix}xyz(x^2-y^2)\times\\(y^2-z^2)(z^2-x^2)\end{matrix}$&$222(R^{}_{1}),223(R^{}_{1}),230(H^{}_{1})$\\ \hline
		\bottomrule
	\end{tabular}
	\label{T:3}
\end{table*}

\begin{table}[h]
	\caption{Tight-binding coefficients, and odd-parity spin splittings for space groups with two atoms per unit cell related by inversion. Abbreviation $c_i\equiv\cos k_i$, $s_i\equiv\sin k_i$, $c_{i/2}\equiv\cos \frac{k_i}{2}$, $s_{i/2}\equiv\sin \frac{k_i}{2}$, $f_x\equiv\sin k_x+\sin\frac{k_x}{2}\cos\frac{\sqrt{3}k_y}{2}$, $f_y\equiv \sqrt{3}\cos\frac{k_x}{2}\sin\frac{\sqrt{3}k_y}{2}$, $f_{x/2}\equiv\sqrt{3}\cos\frac{k_y}{2\sqrt{3}}\sin\frac{k_x}{2}$, $f_{y/2}\equiv \sin \frac{k_y}{\sqrt{3}}+\sin\frac{k_y}{2\sqrt{3}}\cos\frac{k_x}{2}$, and $c_{1/2}\equiv\cos(\frac{k_y}{\sqrt{3}})+2\cos(\frac{k_x}{2})\cos(\frac{k_y}{2\sqrt{3}})$ applies. Wave-vectors $\bf Q$ that do not host 2D irreducible representations are not included. $(s_x,s_z)$ denotes linear combination of $s_x$ and $s_z$. Only leading hopping coefficients are included, while the spin splitting can come from higher-order hopping coefficients. For example, $s_xs_ys_z$ in 11(2e) requires $t_x=c_1c_{y/2}+c_2s_xc_{y/2}s_z$. Trivial factors like $c_x+c_y$ have been omitted. }
	\begin{tabular}{|c|c|c|c|c|c|c|c|c|c|}
		\hline
		SG&$t_x$&$t_y$&$\begin{pmatrix}\pi\\0\\0\end{pmatrix}$&$\begin{pmatrix}0\\ \pi\\0\end{pmatrix}$&$\begin{pmatrix}0\\0\\ \pi\end{pmatrix}$&$\begin{pmatrix}0\\ \pi\\ \pi\end{pmatrix}$&$\begin{pmatrix}\pi\\0\\ \pi\end{pmatrix}$&$\begin{pmatrix}\pi\\ \pi\\0\end{pmatrix}$&$\begin{pmatrix}\pi\\ \pi\\ \pi\end{pmatrix}$\\ \hline
		
		11(2e)&$c_{y/2}$&$c_{y/2}(s_x,s_z)$&&$\begin{matrix}s_y,\\s_xs_ys_z\end{matrix}$&&$\begin{matrix}s_y,\\s_xs_ys_z\end{matrix}$&&$\begin{matrix}s_y,\\s_xs_ys_z\end{matrix}$&$\begin{matrix}s_y,\\s_xs_ys_z\end{matrix}$\\ \hline
		
		13(2e,2f)&$c_{z/2}$&$s_y(c_{z/2},s_xs_{z/2})$&&&$s_x,s_z$&$s_x,s_z$&$s_x,s_z$&&$s_x,s_z$\\ \hline
		
		48(2a-2d)&$c_{x/2}c_{y/2}c_{z/2}$&$s_{x/2}s_{y/2}s_{z/2}$&$s_x$&$s_y$&$s_z$&$s_x$&$s_y$&$s_z$&\\ \hline
		
		49(2e-2h)&$c_{z/2}$&$s_xs_ys_{z/2}$&&&$s_z$&$s_z$&$s_z$&&$s_z$\\ \hline
		
		50(2a-2d)&$c_{x/2}c_{y/2}$&$s_{x/2}s_{y/2}s_z$&$s_x$&$s_y$&&$s_y$&$s_x$&$s_z$&$s_z$\\ \hline
		
		51(2e,2f)&$c_{x/2}$&$c_{x/2}s_z$&$s_x$&&&&$s_x$&$s_x$&$s_x$\\ \hline
		
		59(2a,2b)&$c_{x/2}c_{y/2}$&$c_{x/2}c_{y/2}s_z$&$s_x$&$s_y$&&$s_y$&$s_x$&$s_xs_ys_z$&$s_xs_ys_z$\\ \hline
		
		84(2e,2f)&$c_{z/2}$&$\begin{matrix}s_xs_ys_{z/2},\\(c_x-c_y)s_{z/2}\end{matrix}$&&&$s_z$&$s_z$&$s_z$&&$s_z$\\ \hline
		
		85(2a,2b)&$c_{x/2}c_{y/2}$&$\begin{matrix}s_{x/2}s_{y/2}s_z,\\c_{x/2}c_{y/2}(c_x-c_y)s_z\end{matrix}$
		&$s_x$&$s_y$&&$s_y$&$s_x$&$\begin{matrix}s_z,\\(c_x-c_y)s_xs_ys_z\end{matrix}$&$\begin{matrix}s_z,\\(c_x-c_y)s_xs_ys_z\end{matrix}$\\ \hline
		
		85(2c)&$c_{x/2}c_{y/2}$&$\begin{matrix}c_{x/2}c_{y/2}s_z,\\s_{x/2}s_{y/2}(c_x-c_y)s_z\end{matrix}$
		&$s_x$&$s_y$&&$s_y$&$s_x$&$\begin{matrix}s_xs_ys_z,\\(c_x-c_y)s_z\end{matrix}$&$\begin{matrix}s_xs_ys_z,\\(c_x-c_y)s_z\end{matrix}$\\ \hline
		
		86(2a,2b)&$c_{x/2}c_{y/2}c_{z/2}$&$\begin{matrix}s_{x/2}s_{y/2}s_{z/2},\\c_{x/2}c_{y/2}(c_x-c_y)s_{z/2}\end{matrix}$
		&$s_x$&$s_y$&$s_z$&$s_x,s_y$&$s_x,s_y$&$\begin{matrix}s_z,\\(c_x-c_y)s_xs_ys_z\end{matrix}$&\\ \hline
		
		124(2a,2c)&$c_{z/2}$&$s_xs_y(c_x-c_y)s_{z/2}$&&&$s_z$&$s_z$&$s_z$&&$s_z$\\ \hline
		
		125(2a,2b)&$c_{x/2}c_{y/2}$&$s_{x/2}s_{y/2}(c_x-c_y)s_z$&$s_x$&$s_y$&&$s_y$&$s_x$&$(c_x-c_y)s_z$&$(c_x-c_y)s_z$\\ \hline
		
		125(2c,2d)&$c_{x/2}c_{y/2}$&$s_{x/2}s_{y/2}s_z$&$s_x$&$s_y$&&$s_y$&$s_x$&$s_z$&$s_z$\\ \hline
		
		126(2a,2b)&$c_{x/2}c_{y/2}c_{z/2}$&$s_{x/2}s_{y/2}(c_x-c_y)s_{z/2}$&$s_x$&$s_y$&$s_z$&$s_x$&$s_y$&$(c_x-c_y)s_z$&$s_xs_ys_z$\\ \hline
		
		129(2a,2b)&$c_{x/2}c_{y/2}$&$c_{x/2}c_{y/2}(c_x-c_y)s_z$&$s_x$&$s_y$&&$s_y$&$s_x$&$s_xs_y(c_x-c_y)s_z$&$s_xs_y(c_x-c_y)s_z$\\ \hline
		
		129(2c)&$c_{x/2}c_{y/2}$&$c_{x/2}c_{y/2}s_z$&$s_x$&$s_y$&&$s_y$&$s_x$&$s_xs_ys_z$&$s_xs_ys_z$\\ \hline
		
		131(2e,2f)&$c_{z/2}$&$(c_x-c_y)s_{z/2}$&&&$s_z$&$s_z$&$s_z$&&$s_z$\\ \hline
		
		132(2b,2d)&$c_{z/2}$&$s_xs_ys_{z/2}$&&&$s_z$&$s_z$&$s_z$&&$s_z$\\ \hline
		
		134(2a,2b)&$c_{x/2}c_{y/2}c_{z/2}$&$s_{x/2}s_{y/2}s_{z/2}$&$s_x$&$s_y$&$s_z$&$s_x$&$s_y$&$s_z$&\\ \hline
		
		137(2a,2b)&$c_{x/2}c_{y/2}c_{z/2}$&$c_{x/2}c_{y/2}(c_x-c_y)s_{z/2}$&$s_x$&$s_y$&$s_z$&$s_y$&$s_x$&$s_xs_y(c_x-c_y)s_z$&$s_xs_ys_z$\\ \hline
		
		201(2a)&$c_{x/2}c_{y/2}c_{z/2}$&$s_{x/2}s_{y/2}s_{z/2}$&$s_x$&$s_y$&$s_z$&$s_x$&$s_y$&$s_z$&\\ \hline
		
		222(2a)&$c_{x/2}c_{y/2}c_{z/2}$&$\begin{matrix}s_{x/2}s_{y/2}s_{z/2}(c_y-c_z)\\(c_z-c_x)(c_x-c_y)\end{matrix}$&$s_x$&$s_y$&$s_z$&$\begin{matrix}
			s_x\times\\(c_y-c_z)\end{matrix}$&$\begin{matrix}
			s_y\times\\(c_z-c_x)\end{matrix}$&$\begin{matrix}
			s_z\times\\(c_x-c_y)\end{matrix}$&$s_xs_ys_z$\\ \hline
		
		224(2a)&$c_{x/2}c_{y/2}c_{z/2}$&$s_{x/2}s_{y/2}s_{z/2}$&$s_x$&$s_y$&$s_z$&$s_x$&$s_y$&$s_z$&\\ \hline
	\end{tabular}
	\begin{tabular}{|c|c|c|c|c|c|}
		\hline
		SG&$t_x$&$t_y$&$\begin{pmatrix}0\\0\\ \pi\end{pmatrix}$&$\begin{pmatrix}\pi\\0\\ 0\end{pmatrix}$&$\begin{pmatrix}\pi\\0\\ \pi\end{pmatrix}$\\ \hline
		
		163(2a)&$c_{z/2}$&$f_y(f_y^2-3f_x^2)c_{z/2}$&$s_z$&&$s_z$\\ \hline
		
		163(2c,2d)&$c_{1/2}c_{z/2}$&$f_{y/2}(f_{y/2}^2-3f_{x/2}^2)c_{z/2}$&$s_z$&&$s_z$\\ \hline
		
		165(2a)&$c_{z/2}$&$f_x(f_x^2-3f_y^2)c_{z/2}$&$s_z$&&$s_z$\\ \hline
		
		176(2a)&$c_{z/2}$&$\begin{matrix}f_x(f_x^2-3f_y^2)c_{z/2},\\f_y(f_y^2-3f_x^2)c_{z/2}\end{matrix}$&$s_z$&&$s_z$\\ \hline
		
		176(2c,2d)&$c_{1/2}c_{z/2}$&$\begin{matrix}f_{y/2}(f_{y/2}^2-3f_{x/2}^2)c_{z/2},\\f_{x/2}(f_{x/2}^2-3f_{y/2}^2)c_{z/2}\end{matrix}$&$s_z$&&$s_z$\\ \hline
		
		192(2a)&$c_{z/2}$&$\begin{matrix}f_x(f_x^2-3f_y^2)\times\\f_y(f_y^2-3f_x^2)s_{z/2}\end{matrix}$&$s_z$&&$s_z$\\ \hline
		
		193(2a)&$c_{z/2}$&$f_x(f_x^2-3f_y^2)c_{z/2}$&$s_z$&&$s_z$\\ \hline
		
		194(2b)&$c_{z/2}$&$f_y(f_y^2-3f_x^2)c_{z/2}$&$s_z$&&$s_z$\\ \hline
		
		194(2c,2d)&$c_{1/2}c_{z/2}$&$f_{y/2}(f_{y/2}^2-3f_{x/2}^2)c_{z/2}$&$s_z$&&$s_z$\\ \hline
	\end{tabular}
	\label{T:4}
\end{table}

\begin{table}[h]
	\caption{Tight-binding coefficients, and odd-parity spin splittings for space groups with two atoms per unit cell at inversion center.  The same abbreviation applies.}
	\begin{tabular}{|c|c|c|c|c|c|c|c|c|c|}
		\hline
		SG&$t_x$&$t_z$&$\begin{pmatrix}\pi\\0\\0\end{pmatrix}$&$\begin{pmatrix}0\\ \pi\\0\end{pmatrix}$&$\begin{pmatrix}0\\0\\ \pi\end{pmatrix}$&$\begin{pmatrix}0\\ \pi\\ \pi\end{pmatrix}$&$\begin{pmatrix}\pi\\0\\ \pi\end{pmatrix}$&$\begin{pmatrix}\pi\\ \pi\\0\end{pmatrix}$&$\begin{pmatrix}\pi\\ \pi\\ \pi\end{pmatrix}$\\ \hline
		
		11(2a-2d)&$c_{y/2}$&$s_y(s_x,s_z)$&&$s_y$&&$s_y$&&$s_y$&$s_y$\\ \hline
		
		13(2a-2d)&$c_{z/2},s_xs_{z/2}$&$s_y(s_x,s_z)$&&&$s_z$&$s_z$&$s_z$&&$s_z$\\ \hline
		
		14(2a-2d)&$c_{y/2}(s_xs_{z/2},c_{z/2})$&$s_y(s_x,s_z)$&&$s_y$&$s_z$&&$s_z$&$s_y$&\\ \hline
		
		49(2a-2d)&$c_{z/2}$&$s_xs_y$&&&$s_z$&$s_z$&$s_z$&&$s_z$\\ \hline
		
		51(2a-2d)&$c_{x/2}$&$s_xs_z$&$s_x$&&&&$s_x$&$s_x$&$s_x$\\ \hline
		
		53(2a-2d)&$c_{x/2}c_{z/2}$&$s_ys_z$&$s_x$&&$s_z$&$s_z$&&$s_x$&\\ \hline
		
		55(2a-2d)&$c_{x/2}c_{y/2}$&$s_xs_y$&$s_x$&$s_y$&&$s_y$&$s_x$&&\\ \hline
		
		58(2a-2d)&$c_{x/2}c_{y/2}c_{z/2}$&$s_xs_y$&$s_x$&$s_y$&$s_z$&&&&$s_xs_ys_z$ \\ \hline

		84(2a,2b)&$c_{z/2}$&$(c_x-c_y),s_xs_y$&&&$s_z$&&&&$s_z$\\ \hline
		
		84(2c,2d)&$\begin{matrix}c_{x/2}c_{y/2}c_{z/2},\\s_{x/2}s_{y/2}c_{z/2}\end{matrix}
		$&$(c_x-c_y),s_xs_y$&&&$s_z$&&&&$s_xs_ys_z$\\ \hline
		
		124(2b,2d)&$c_{z/2}$&$s_xs_y(c_x-c_y)$&&&$s_z$&$s_z$&$s_z$&&$s_z$\\ \hline
		
		127(2a,2b)&$c_{x/2}c_{y/2}$&$s_xs_y(c_x-c_y)$&$s_x$&$s_y$&&$s_y$&$s_x$&&\\ \hline
		
		127(2c,2d)&$c_{x/2}c_{y/2}$&$s_xs_y$&$s_x$&$s_y$&&$s_y$&$s_x$&&\\ \hline
		
		128(2a,2b)&$c_{x/2}c_{y/2}c_{z/2}$&$s_xs_y(c_x-c_y)$&$s_x$&$s_y$&$s_z$&&&&$s_xs_ys_z$ \\ \hline
		
		131(2a,2b)&$c_{z/2}$&$(c_x-c_y)$&&&$s_z$&&&&$s_z$\\ \hline
		
		131(2c,2d)&$c_{x/2}c_{y/2}c_{z/2}$&$(c_x-c_y)$&&&$s_z$&&&&$s_xs_ys_z$\\ \hline
		
		132(2a,2c)&$c_{z/2}$&$s_xs_y$&&&$s_z$&$s_z$&$s_z$&&$s_z$\\ \hline
		
		136(2a,2b)&$c_{x/2}c_{y/2}c_{z/2}$&$s_xs_y$&$s_x$&$s_y$&$s_z$&&&&$s_xs_ys_z$ \\ \hline
		
		223(2a)&$c_{x/2}c_{y/2}c_{z/2}$&$(c_x-c_y)(c_y-c_z)(c_z-c_x)$&$s_x$&$s_y$&$s_z$&&&&$s_xs_ys_z$ \\ \hline
		
	\end{tabular}
	
	\begin{tabular}{|c|c|c|c|c|c|}
		\hline
		SG&$t_x$&$t_z$&$\begin{pmatrix}0\\0\\ \pi\end{pmatrix}$&$\begin{pmatrix}\pi\\0\\ 0\end{pmatrix}$&$\begin{pmatrix}\pi\\0\\ \pi\end{pmatrix}$\\ \hline
		
		163(2b)&$\begin{matrix} c_{z/2},\\f_x(3f_y^2-f_x^2)s_{z/2}\end
			{matrix}$&$\begin{matrix} f_y(f_y^2-3f_x^2)s_z,\\f_xf_y(f_x^2-3f_y^2)(3f_x^2-f_y^2)\end
				{matrix}$&$s_z$&&$s_z$\\ \hline
				
				165(2b)&$\begin{matrix} c_{z/2},\\f_y(3f_x^2-f_y^2)s_{z/2}\end
					{matrix}$&$\begin{matrix} f_x(f_x^2-3f_y^2)s_z,\\f_xf_y(f_x^2-3f_y^2)(3f_x^2-f_y^2)\end
						{matrix}$&$s_z$&&$s_z$\\ \hline
						
						176(2b)&$c_{z/2}$&$\begin{matrix}s_zf_x(f_x^2-3f_y^2),\\s_zf_y(f_y^2-3f_x^2)\end{matrix}$&$s_z$&&$s_z$\\ \hline
						
						192(2b)&$c_{z/2}$&$f_xf_y(f_x^2-3f_y^2)(3f_x^2-f_y^2)$&$s_z$ &&$s_z$\\ \hline
						
						193(2b)&$c_{z/2}$&$s_zf_x(f_x^2-3f_y^2)$&$s_z$ & &$s_z$ \\ \hline
						
						194(2a)&$c_{z/2}$&$s_zf_y(f_y^2-3f_x^2)$&$s_z$&&$s_z$ \\ \hline
					\end{tabular}
					\label{T:5}
				\end{table}

				\section{Electronic structure of FeSe}
				Density-functional theory calculations for bulk FeSe with the $(\pi,\pi,0)$ coplanar magnetic order are carried out using the Full-potential Linearized Augmented Plane Wave (FLAPW) method
				\cite{FLAPW2009}, with the scalar-relativistic approximation \cite{KH1977,MPK1980}.
				The Perdew-Burke-Ernzerhof form of exchange correlation functional \cite{PBE} is used, along with wave function and potential energy cutoffs of 16 and 200 Ry, respectively. 
				A muffin-tin sphere radius of 1.1 \AA{} is applied to both Fe and Se atoms.
				The superlattice setup is $\sqrt{2}\times\sqrt{2}\times 1$, as indicated by the dotted line in the left panel of Figure 2 of the main text, with experimental lattice parameters ($a=3.765$ \AA; $c=5.518$ \AA; $z_\textrm{Se}=0.26$) \cite{pearson1997handbook}.  The Brillouin zone is sampled using (15,15,10) and (70,70,50) k-point meshes for self-consistent field calculation and Fermi surface calculation, respectively.  The resulting band structure agrees with a previous calculation employing the generalized Bloch theorem for spin spirals \cite{TS2018}.

				\section{Electronic structure of CeNiAsO}
				
\begin{figure}[tb]
					\centering
					\includegraphics[width=0.7\linewidth]{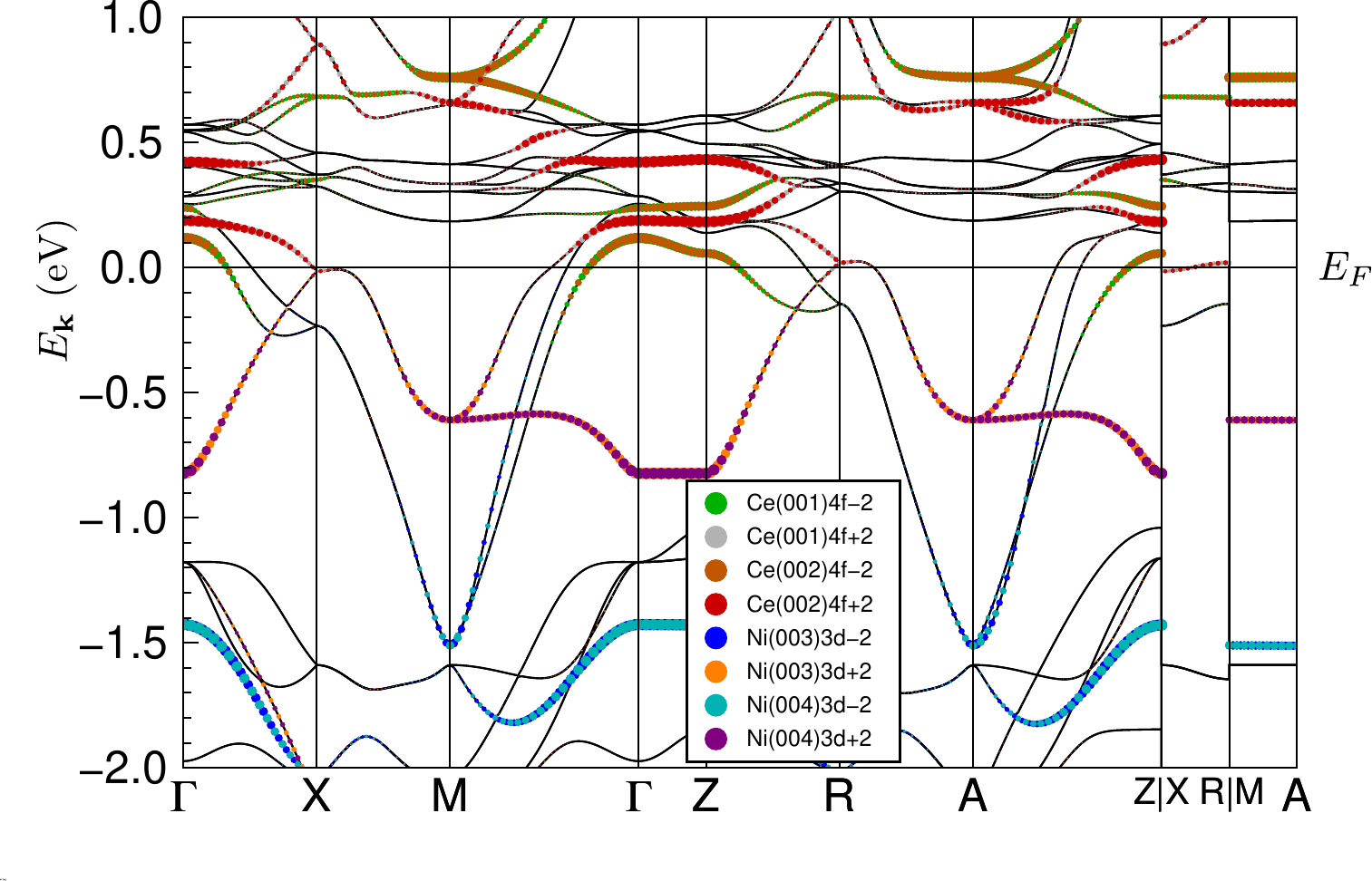}
					\caption{Electronic structure of CeNiAsO including orbital projections to Ce and Ni states. Two pairs of bands cross the Fermi level $E_F$ such that two single band models are relevant to describe the low-energy electronic structure.}
					\label{fig_sm_CeNiAsO}
				\end{figure}

				For the construction of the minimal low-energy models relevant for CeNiAsO, we perform {\it ab initio} calculations within the full-potential local-orbital (FPLO) code \cite{Koepernik1999} (version 22.00-62) in the non-relativistic setting and the LDSA approximation. We use $a=b=4.014537 \mathrm{\AA}$, $c=7.90723\mathrm{\AA}$ in space group 129, Ce on 2c with internal coordinates $(1/4,1/4,0.154739)$, Ni on 2b with $(1/4,1/4,1/2)$, As on 2c with $(1/4,1/4,-0.351368)$ and O on 2a with $(1/4, 3/4,0)$. The resulting band structure is shown in Fig.~\ref{fig_sm_CeNiAsO} which exhibits two pairs of not hybridized bands as checked by calculating the irreps of the corresponding bands.
				
\end{document}